\documentclass[twocolumn,trackchanges]{aastex63}
\usepackage{amsmath}
\usepackage{inputenc}
\makeatletter
\newcommand{\pushright}[1]{\ifmeasuring@#1\else\omit\hfill$\displaystyle#1$\fi\ignorespaces}
\newcommand{\pushleft}[1]{\ifmeasuring@#1\else\omit$\displaystyle#1$\hfill\fi\ignorespaces}
\makeatother

\received{Apr. 20, 2021}
\revised{June 20, 2021}
\accepted{June 22, 2021}
\published{Sep. 21, 2021}
\submitjournal{PSJ}

\shorttitle{Oscillations in differentially rotating models of Saturn}
\shortauthors{Dewberry et al.}
\graphicspath{{./}{figures/}}
\begin{document}

\title{Constraining Saturn's interior with ring seismology: effects of differential rotation and stable stratification}

\correspondingauthor{Janosz W. Dewberry}
\email{jdewberry@cita.utoronto.ca}

\author[0000-0001-9420-5194]{Janosz W. Dewberry}
\affiliation{Canadian Institute for Theoretical Astrophysics, 60 St. George Street, Toronto, ON M5S 3H8, Canada}

\author[0000-0002-4940-9929]{Christopher R. Mankovich}
\affiliation{Division of Geological and Planetary Sciences, California Institute of Technology, Pasadena, CA 91125, USA}

\author[0000-0002-4544-0750]{Jim Fuller}
\affiliation{TAPIR, Walter Burke Institute for Theoretical Physics, Mailcode 350-17, California Institute of Technology, Pasadena, CA 91125, USA}

\author[0000-0002-1934-6250]{Dong Lai}
\affiliation{Cornell Center for Astrophysics and Planetary Science, Department of Astronomy, Cornell University, Ithaca, NY 14853, USA}
\affiliation{Tsung-Dao Lee Institute, Shanghai Jiao Tong University, Shanghai 200240, People's Republic of China}

\author[0000-0002-9408-2857]{Wenrui Xu}
\affiliation{Department of Astrophysical Sciences, Princeton University, Peyton Hall, Princeton, NJ 08544, USA}



\begin{abstract}
Normal mode oscillations in Saturn excite density and bending waves in the C Ring, providing a valuable window into the planet's interior. Saturn's fundamental modes (f~modes) excite the majority of the observed waves, while gravito-inertial modes (rotationally modified g~modes) associated with stable stratification in the deep interior provide a compelling explanation for additional density waves with low azimuthal wavenumbers $m.$ However, multiplets of density waves with nearly degenerate frequencies, including an $m=3$ triplet, still lack a definitive explanation. We investigate the effects of rapid and differential rotation on Saturn's oscillations, calculating normal modes for independently constrained interior models. We use a non-perturbative treatment of rotation that captures the full effects of the Coriolis and centrifugal forces, and consequently the mixing of sectoral f~modes with g~modes characterized by very different spherical harmonic degrees. Realistic profiles for differential rotation associated with Saturn's zonal winds can enhance these mode interactions, producing detectable oscillations with frequencies separated by less than $1\%$. Our calculations demonstrate that a three-mode interaction involving an f~mode and two g~modes can feasibly explain the finely split $m=3$ triplet, although the fine-tuning required to produce such an interaction generally worsens agreement with seismological constraints provided by $m=2$ density waves. Our calculations additionally demonstrate that sectoral f~mode frequencies are measurably sensitive to differential rotation in Saturn's convective envelope. Finally, we find that including realistic equatorial antisymmetry in Saturn's differential rotation profile couples modes with even and odd equatorial parity, producing oscillations that could in principle excite both density and bending waves simultaneously.
\end{abstract}



\section{Introduction}\label{sec:intro}
The interiors of gas giants remain enigmatic, even within our own solar system. In particular, we still know very little about the extent and composition of Saturn's core. \textit{Cassini}'s Grand Finale facilitated precise measurement of the planet's non-spherical gravitational field \citep{Iess2019}, but even knowledge of high-order harmonic coefficients (through $J_{12}$) leaves room for significant degeneracy in deep-interior density profiles \citep{Movshovitz2020}. Uncertainties surrounding the equation of state for hydrogen and helium mixtures at pressures beyond the reach of laboratory experiments, not to mention the concentration and dissolution of heavier elements, complicate the process of bridging this observational gap \citep[e.g.,][]{Helled2018}.

Fortunately, Saturn's rings present a rare opportunity to place seismological constraints on the planet's interior. Analyzing occultation data obtained by \emph{Cassini}, \citet{Hedman2013} validated a decades-old prediction \citep{Marley1991,Marley1993} that Saturn's ``fundamental'' oscillation modes (f modes) excite density waves at Lindblad resonances in the C Ring. Continued analysis of \emph{Cassini} observations has revealed a wealth of additional density (and bending) waves, thought to be excited by planetary oscillations with even (odd) equatorial symmetry \citep{Hedman2014,French2016,French2019,Hedman2019}. 

Saturn ring seismology has already provided insight into the planet's internal structure. Recently, \citet{Mankovich2019} used f mode identifications to measure Saturn's mean internal rotation rate to a high degree of accuracy \citep[a task previously made difficult by the axisymmetry and nearly exact polar alignment of Saturn's magnetic field; ][]{Cao2020}. One of the earliest surprises to come out of Saturn ring seismology was a strong indication that the planet is not, as historically thought, composed simply of a solid core and a convective envelope \citep[but see, e.g., ][]{Leconte2013}: \citet{Fuller2014b} showed that observations of multiple density waves with the same azimuthal wavenumber $m$ and nearly degenerate frequencies could be explained by the rotational mixing of f modes with gravito-inertial modes (hereafter g modes) associated with stable stratification in the deep interior. 

Based on this analysis, \citet{Fuller2014b} predicted the appearance of additional density waves driven by low-radial order g modes. \citet{French2016,French2019} subsequently observed density waves identifiable with such g modes, which \citet{Mankovich2021} recently fit jointly with Saturn's gravity field to find that Saturn's core is likely very diffuse, and the region of stable stratification associated with composition gradients very extensive. However, the calculations of \citet{Fuller2014b} and \citet{Mankovich2021} encounter difficulty in reproducing the finest frequency splittings of less than $1\%$ observed for a triplet of $m=3$ density waves \citep{Hedman2013}. This difficulty could be related to the authors' method for including the effects of rotation, which they approximated as rigid-body and treated using second-order perturbation theory. 

In this paper, we revisit both of these approximations. We first calculate oscillations from the rigidly rotating Saturn models of \citet{Mankovich2021} without making any approximation with respect to the modes' modification by either the Coriolis force or the centrifugal distortion of the equilibrium planetary structure. We then investigate the additional coupling of these pulsations by differential rotation associated with Saturn's zonal winds. 

Along with conventional (but rotationally modified) g modes, these calculations reveal sequences of so-called ``rosette'' modes \citep{Ballot2012,Takata2013} that can only be described with higher-order treatments of rotation. Like the relatively low-degree g modes involved in the rotational mixing considered by \citet{Fuller2014b}, both rosette modes and high-degree g modes gain enhanced surface gravitational perturbations from avoided crossings with f modes. The density of the high-$\ell$ g mode spectrum in frequency-space presents numerous opportunities for such mixing, but the avoided crossings are proportionately narrower. Rigidly rotating models therefore require fine-tuning to produce multiple detectable modes with finely split frequencies.

Limited to the outer envelope, differential rotation from Saturn's zonal winds primarily affects the f modes in our calculations. The zonal winds produce small, but significant shifts in the frequencies of high-degree sectoral ($\ell\sim m$) f modes. Varying the depth of wind decay, we find that differential rotation can enhance rotational mixing between the low-$\ell$ sectoral f modes and the underlying dense spectrum of high-degree g modes. We present examples of three-mode interactions that could in principle explain the observed $m=3$ triplet of density waves with finely split frequencies, although producing these interactions still requires fine-tuning (which generally worsens the agreement of our calculations of other low-order modes with observations). 

Finally, we find that realistic equatorial antisymmetry in the differential rotation profile couples oscillation modes with even and odd equatorial parity. This coupling is weak, but sufficient to produce equatorially \emph{asymmetric} oscillations that could simultaneously excite density and bending waves at separate locations in the rings. Observational confirmation of such simultaneous density/bending wave excitation would provide a very useful constraint on Saturn's interior, since the weak antisymmetric component of Saturn's zonal winds is likely only capable of mixing f modes with a limited subset of opposite-parity g modes of low degree and low order.

This paper takes the following structure: \autoref{sec:models} and \autoref{sec:pulse} introduce our models of Saturn and our method for calculating their associated oscillation modes, respectively. \autoref{sec:rig} and \autoref{sec:diff} then present the results of our calculations with purely rigid and differential rotation (resp.), \autoref{sec:disc} provides discussion, and \autoref{sec:conc} summarizes our conclusions. The Appendices provide further details about our numerical methods.

\newpage
\section{Planetary models}\label{sec:models}
\subsection{Rigidly rotating models}\label{sec:rmod}
We start from the rigidly rotating, oblate models of \citet{Mankovich2021}. Calculated using a fourth-order theory of figures \citep{Nettelmann2017}, these parameterize Saturn's interior structure via a smooth variation in heavy element and Helium mass fractions over a core-envelope transition region. The models are thermally adiabatically stratified and adopt a temperature of 135 K at $P=1$ bar. We assume an equatorial radius $R_S=60.268\times10^3$km, $GM_S=37,931,207.7\text{km}^3\text{s}^{-2}$ \citep{Jacobson2006}, 
and a $10.561$hr period \citep{Mankovich2019} corresponding to a bulk rotation rate $\Omega_S\simeq0.397\Omega_\text{dyn}$, where $\Omega_\text{dyn}=\sqrt{GM_S/R_S^3}$ is the equatorial Keplerian rotation rate.

\begin{figure}
    \centering
    \includegraphics[width=\columnwidth]{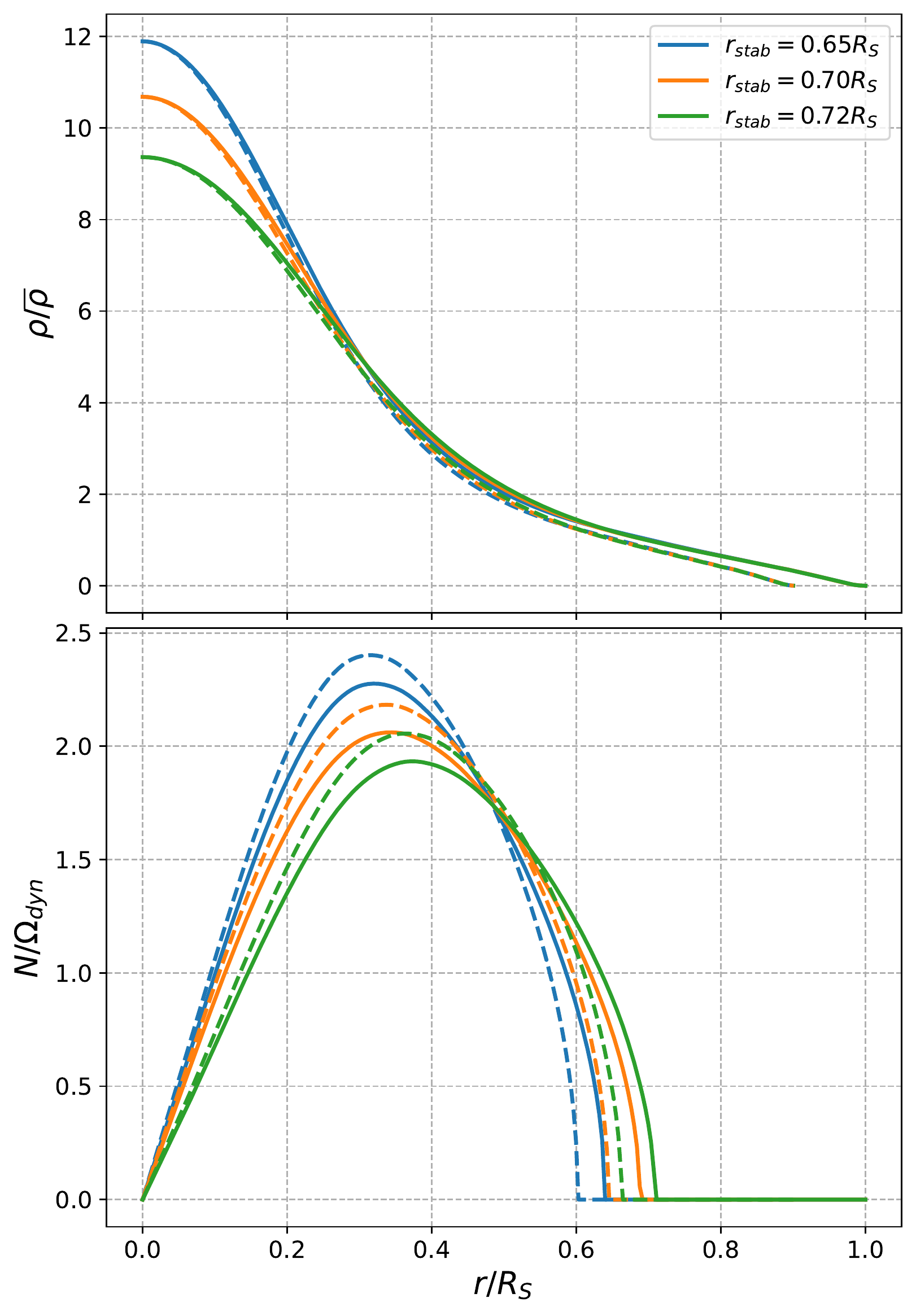}
    \caption{Plots showing equatorial (solid) and polar (dashed) profiles of density (top) and Brunt-V\"ais\"al\"a frequency (bottom) for the three models considered in this paper. Density is normalized by $\overline{\rho}:=M/(4\pi R_S^3/3)$, while the buoyancy frequency is given in units of $\Omega_\text{dyn}=\sqrt{GM_S/R_S^3}$.}
    \label{fig:models}
\end{figure}

\autoref{fig:models} plots profiles of density (top) and Brunt-V\"ais\"al\"a frequency (bottom) for the models considered in this paper, which have stably stratified regions extending from $r=0$ to $r_\text{stab}=0.65R_S,0.70R_S,$ and $0.72R_S$ at the equator. We concentrate on these three because they illustrate different aspects of normal mode coupling by differential rotation, and take the model with $r_\text{stab}=0.70R_S$ as fiducial because it agrees most closely with the best-fit found by \citet{Mankovich2021} for smooth composition gradients. This work focuses on exploring the potential effects of differential rotation on oscillations in Saturn, not providing an exact description for the planet's interior. However, we do match the rigid-body component of the observed zonal gravity harmonic $J_2$ to within $1\%$, and $J_4,J_6,J_8$ to within $5\%$ \citep{Iess2019,Galanti2021}. 

\subsection{Zonal winds}\label{sec:zwind}
\begin{figure*}
    \centering
    \includegraphics[width=0.7\textwidth]{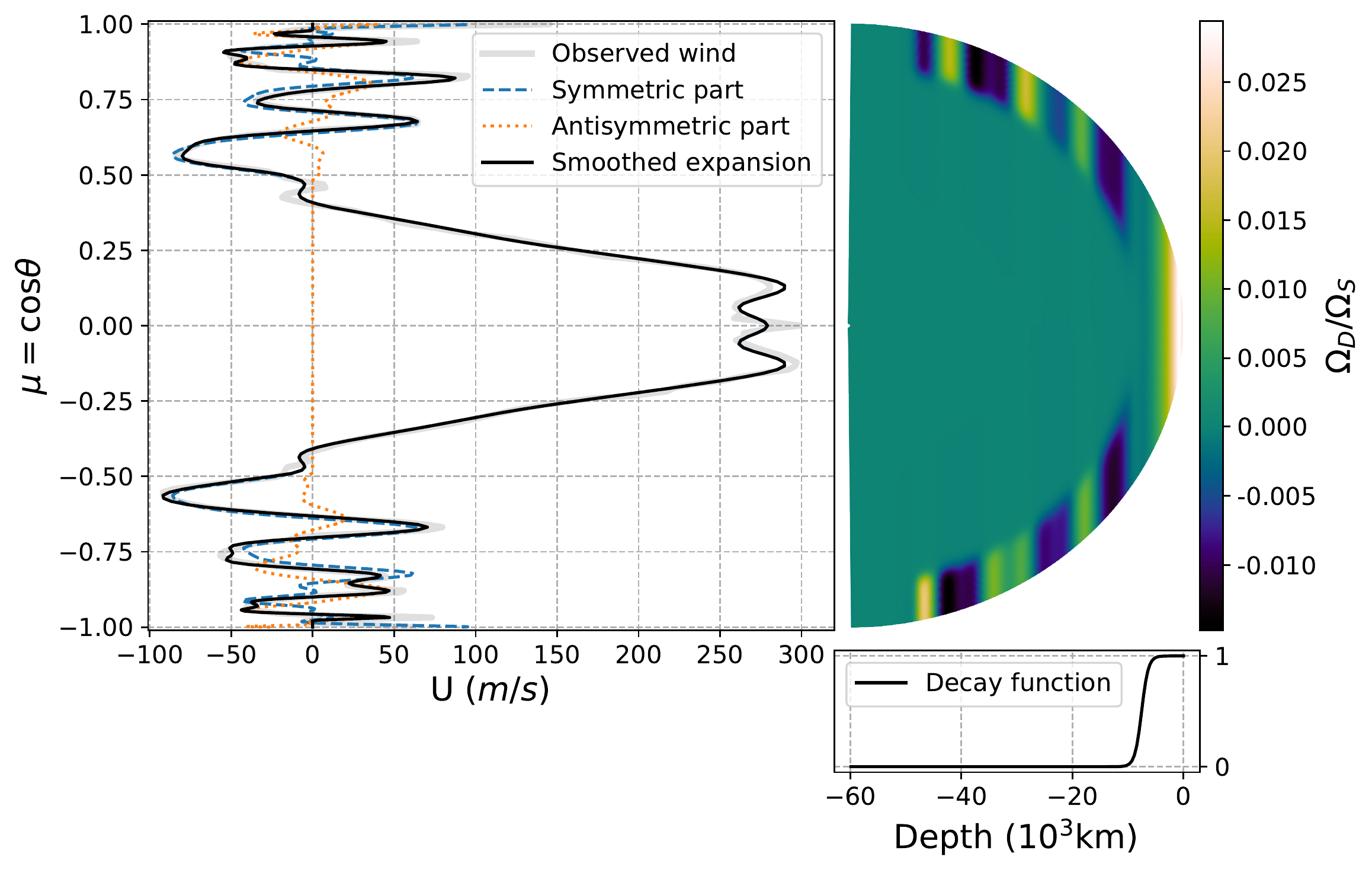}
    \caption{Top left: Saturn's observed zonal wind (gray line), superimposed by symmetric (blue dashes), antisymmetric (orange dots) and full (black line) expansions in Legendre polynomials. Top right: differential rotation profile (normalized by bulk rotation rate) produced by this zonal wind and decay profile. Bottom right: decay profile with $d=0.125R_S$. }
    \label{fig:satWind}
\end{figure*}

Our oscillation calculations directly incorporate Saturn's observed wind profiles, as reported by \citet{Garcia-Melendo2011} and adjusted for the rotation period of $10.561$hr constrained by \citet{Mankovich2019}. \autoref{fig:satWind} (top left) plots the zonal wind speed with $\mu=\cos\theta$ (here $\theta$ is the colatitude), which we smooth and relax to zero at the poles by expanding the surface azimuthal velocity as $U=w\sum_{l=0}^{60}c_{l}P_{l}(\mu)$. Here $c_{l}$ are coefficients calculated by projecting onto Legendre polynomials $P_l,$ and $w$ is a window function that transitions smoothly to zero for $|\mu|>0.975$. 

Informed by recent investigations combining magnetic and gravity measurements \citep{Galanti2021}, we assume that the zonal wind penetrates deeply as barotropic ``rotation on cylinders,'' producing a perturbation to the angular velocity with the form $R\Omega_D=\eta U.$ Here  $R=r\sin\theta$ is the distance to the rotation axis (cylindrical radius), and $\eta$ defines a decay function. We adopt the simplified relation
\begin{equation}
    \eta(r,\mu)=\frac{1}{2}\left\{1 + \tanh\left[\frac{1}{\Delta}(r-r_s+d)\right]\right\},
\end{equation} 
where $r_s(\mu)$ tracks the oblate planetary surface, while $\Delta$ and $d$ parameterize the width and depth of decay to rigid rotation in the interior. We take $\Delta=0.02R_s,$ and vary the decay depth. \autoref{fig:satWind} (bottom) plots an example decay function with a fiducial value of $d=0.125R_S\simeq7.5\times10^3$km, which produces an angular velocity perturbation $\Omega_D$ with the meridional profile shown in \autoref{fig:satWind} (top right), and a maximum deviation from the bulk rotation rate of $\lesssim 3\%$ at the equator. 

As demonstrated by the orange dotted line in \autoref{fig:satWind} (top left), Saturn's zonal winds possess an antisymmetric component that is small, but nonzero at high latitudes. This antisymmetric component can weakly couple modes with even and odd equatorial parity (even and odd $\ell-m$), which would otherwise be unable to interact (see \autoref{sec:diff}). Unless otherwise stated, we therefore include odd-degree Legendre polynomials in our smoothing expansion, but we note that antisymmetric surface-level winds introduce discontinuities when allowed to extend barotropically to equatorial radii $r(\mu=0)<R_S$. We eliminate these discontinuities in $\Omega_D$ (and more importantly its gradient) by smoothly transitioning the antisymmetric component of the surface wind to zero within $|\mu|<0.5$, using a window function composed of tanh functions with a scale length set to half the polar grid size. The antisymmetric component of the wind is already small for $|\mu|\lesssim0.5$, and our results are insensitive to this choice of window function.

\section{Oscillation calculations}\label{sec:pulse}
This section outlines our methods for calculating oscillation modes. We use a combination of  non-perturbative (\autoref{sec:npert}) and perturbative (\autoref{sec:dmix}) techniques for including the effects of rigid and differential rotation (resp.), since Saturn's bulk rotation rate of $\Omega_S\simeq0.397\Omega_\text{dyn}$ is rapid, but the differential rotation associated with zonal winds constitutes a relatively small perturbation. Throughout this section, we refer to both spherical and cylindrical polar coordinate systems $(r,\theta,\phi)$ and $(R,\phi,z).$

\subsection{Non-perturbative calculations with rigid rotation}\label{sec:npert}
In a frame rotating with constant angular velocity ${\bf\Omega_S}=\Omega_S\hat{\bf z},$ the equations governing a self-gravitating, inviscid gas characterized by velocity ${\bf u}$, pressure $P$, density $\rho$ and gravitational field $\Phi$ are \citep[e.g.,][]{Lynden-Bell1967}
\begin{align}\label{eq:EoM0}
    D_t{\bf u}
    +2{\bf \Omega_S\times u}
    +{\bf \Omega_S\times}({\bf \Omega_S\times r})
    &=-\nabla P/\rho
    -\nabla\Phi,
\\
    D_t\rho&=-\rho\nabla\cdot {\bf u},
\\\label{eq:energy}
    D_tP&=-\Gamma_1P\nabla\cdot {\bf u},
\\
    \nabla^2\Phi&=4\pi G\rho,
\end{align}
where $D_t=\partial_t+{\bf u\cdot\nabla}$ denotes the convective derivative, and $\Gamma_1$ is the first adiabatic exponent. Here we have neglected non-adiabatic heating in the thermal energy equation (\autoref{eq:energy}).

Perturbing these equations around a rigidly rotating, barotropic equilibrium state with pressure $P_0$ and density $\rho_0$ at rest in the rotating frame, the linearized equations governing adiabatic oscillations with the harmonic time dependence $\exp[-\text{i}\omega t]$, where $\omega$ is a (rotating frame) frequency, can be written as
\begin{align}\label{eq:lineq1}
    \text{i}\omega{\bf v}
    &=2{\bf\Omega_S}\times{\bf v}
    -{\bf G}b
    +\rho_0^{-1}\nabla(\rho_0h)
    +\nabla\Phi',
\\
    \text{i}\omega \rho_0 b
    &=\nabla\cdot(\rho_0{\bf v}),
\\
    \text{i}\omega(h-c_A^2b)
    &={\bf v}\cdot({\bf G}-c_A^2\nabla\ln\rho_0),
\\\label{eq:lineq4}
    \nabla^2\Phi'&=4\pi G\rho_0 b.
\end{align}
Here ${\bf G}=\rho_0^{-1}\nabla P_0$ is the effective gravity (including both gravitational and centrifugal accelerations), $c_A^2=\Gamma_1P_0/\rho_0$ is the square of the adiabatic sound speed, and ${\bf v}$ is an Eulerian velocity perturbation. Lastly, with primes denoting Eulerian perturbations to the pressure $P',$ density $\rho'$ and gravitational field $\Phi'$, we have defined the variables $h=P'/\rho_0$ and $b=\rho'/\rho_0.$ 

Saturn's rapid bulk rotation breaks the spherical symmetry of the oscillation equations, through both the introduction of the Coriolis force $2{\bf\Omega_S}\times{\bf v}$, and a centrifugal flattening of the equilibrium state. To account for these effects, both of which can be important for the low-order f-modes and g-modes of interest, we use a non-perturbative numerical method \citep[e.g.,][]{Lignieres2006,Reese2006,Reese2009,Reese2013,Ouazzani2012} that treats the eigenvalue problem defined by Equations \eqref{eq:lineq1}-\eqref{eq:lineq4} as fundamentally two-dimensional. 

In this approach, the perturbations are expanded in a \emph{series} of spherical harmonics and their angular derivatives. For example, we expand a given scalar perturbation $X'$ as 
\begin{equation}
    X'(r,\theta,\phi)
    =\sum_{\ell'=m}^\infty X^{\ell'}(\zeta)Y_{\ell'}^m(\theta,\phi),
\end{equation}
where $\zeta$ is a quasi-radial coordinate that matches the oblate surface of the planetary model at $\zeta=1$, and transitions to spherical radius at $\zeta=0$ and $\zeta=2$. We only solve Laplace's equation in the vacuum region $\zeta\in[1,2]$, which is important to include in the computational domain because it permits the application of boundary conditions for the gravitational potential on a spherical surface \citep{Reese2006}.

Substituting such expansions into the linearized equations and projecting onto an arbitrary spherical harmonic $Y_\ell^m$ (this amounts to a Galerkin spectral treatment of the partial differential equations in the angular direction), the dependence of the background state on both radius and colatitude prevents the clean separation of variables that occurs for non-rotating stellar models. Instead, we are left with an infinite series of coupled ordinary differential equations in $\zeta$ that must be truncated and solved simultaneously. We employ pseudospectral `collocation' \citep[e.g.,][]{Boyd2001} for these calculations (see \autoref{app:npert} for further details). The new code we have developed is similar to that of \citet{Xu2017}, but solves the perturbation equations in a form better-suited to our (non-polytropic) models.

Following \citet{dahlen_tromp_1998}, we normalize the eigenmodes of our rigidly rotating models according to 
\begin{equation}\label{eq:norm}
    \langle \boldsymbol{\xi}_i,\boldsymbol{\xi}_j\rangle
    +(\omega_i+\omega_j)^{-1}
    \langle 
        \boldsymbol{\xi}_i,
        2\text{i}{\bf \Omega_S}\times\boldsymbol{\xi}_j
    \rangle
    =\delta_{ij}.
\end{equation}
Here $\boldsymbol{\xi}=(-\text{i}\omega)^{-1}{\bf v}$ is the Lagrangian displacement assuming a rigidly rotating background state, and the inner product is defined as
$\langle \boldsymbol{\xi}_i,\boldsymbol{\xi}_j\rangle
:=\int_V\rho_0
\boldsymbol{\xi}_i^*\cdot\boldsymbol{\xi}_j
\text{d}V$. 
Under this normalization, rotating-frame mode energies are given simply by $\varepsilon_i=2\omega_i^2$ \citep[e.g.,][]{Schenk2002}.

\newpage
\subsection{Perturbative treatment of differential rotation}\label{sec:dmix}
Defining the Lagrangian displacement more generally as $\boldsymbol{\xi}=\Delta{\bf r}$, we note that Eulerian and Lagrangian perturbations $X'$ and $\Delta X$ are related by $X'=\Delta X-\boldsymbol{\xi}\cdot\nabla X.$ Taking the Lagrangian variation of \autoref{eq:EoM0} then produces \citep[e.g.,][]{Lynden-Bell1967}
\begin{align}\notag
    \tilde{D}_t^2\boldsymbol{\xi}
    +2{\bf \Omega_S\times }\tilde{D}_t\boldsymbol{\xi}
    &=-\tilde{\rho}_0^{-1}\left(
        \nabla P'
        -\tilde{\bf G}\rho'
    \right)
    -\nabla\Phi'
    \\&\hspace{-4em}\label{eq:EoM1}
    +\boldsymbol{\xi}\cdot\nabla
    \left(
        {\bf u}_0\cdot\nabla{\bf u}_0
        +2{\bf \Omega_S\times} {\bf u}_0
    \right),
\end{align}
where $\tilde{D}_t=\partial_t+{\bf u}_0\cdot\nabla$, with ${\bf u}_0$ a velocity field satisfying the mechanical equilibrium
\begin{equation}\label{eq:MechEqm}
    {\bf u}_0\cdot\nabla{\bf u}_0
    +2{\bf \Omega_S\times}{\bf u}_0
    +{\bf \Omega_S\times}({\bf \Omega_S\times r})
    =-\tilde{\bf G}
    -\nabla\tilde{\Phi}_0.
\end{equation}
Here $\tilde{\rho}_0$, $\tilde{\bf G}=\nabla\tilde{P}_0/\tilde{\rho}_0$ and $\tilde{\Phi}_0$ denote a new density, effective gravity, and gravitational field associated with steady flow in the rotating frame; via \autoref{eq:MechEqm}, a non-zero velocity field ${\bf u}_0$ will modify the equilibrium pressure, density and gravity field from those of a rigidly rotating fluid body.

For our application to Saturn, we consider a velocity field with the form ${\bf u}_0={\bf \Omega_D\times r}$, where 
${\bf \Omega_D}=\Omega_D(r,\theta)\hat{\bf z}$ is axially symmetric. Then ${\bf u}_0\cdot\nabla {\bf u}_0=-R\Omega_D^2\hat{\bf R}$, and $2{\bf \Omega_S\times u}_0=-R\Omega_S\Omega_D\hat{\bf R}$. \autoref{fig:satWind} demonstrates that $\Omega_D$ is small-amplitude, but varies rapidly over relatively small spatial scales. We therefore ignore the effect of differential rotation on the background pressure, density and gravitational field in \autoref{eq:MechEqm}, but retain terms involving ${\bf u}_0$ and its gradient in \autoref{eq:EoM1}. 

With such a velocity field and the modal dependence 
$\boldsymbol{\xi}\propto\exp[\text{i}(m\phi-\omega t)]$ (for which prograde modes have positive azimuthal wavenumber $m$ and frequency $\omega$), \autoref{eq:EoM1} can be written in the operator form
\begin{equation}\label{eq:master}
    \omega^2\mathcal{T}[\boldsymbol{\xi}]
    +2\omega \mathcal{W}[\boldsymbol{\xi}]
    +\mathcal{V}[\boldsymbol{\xi}]
    +\mathcal{U}[\boldsymbol{\xi}]=0,
\end{equation}
where $\mathcal{T}$ is the identity (represented with this notation because $\langle \boldsymbol{\xi},\mathcal{T}[\boldsymbol{\xi}]\rangle$ relates to kinetic energy), and 
\begin{align}
    \mathcal{W}[\boldsymbol{\xi}]
    &=\text{i}({\bf \Omega_S}+{\bf \Omega_D})\times \boldsymbol{\xi}
    -m\Omega_D\boldsymbol{\xi},
\\
    \mathcal{V}[\boldsymbol{\xi}]
    &={\bf G}b-\rho_0^{-1}\nabla(\rho_0 h)
    -\nabla\Phi',
\\\label{eq:Ueq1}
    \mathcal{U}[\boldsymbol{\xi}]
    &=m^2\Omega_D^2\boldsymbol{\xi}
    -2\text{i} m\Omega_D({\bf \Omega_S+\Omega_D}) 
    \times\boldsymbol{\xi}
    \\\notag&\pushright{
        -2R(\Omega_S+\Omega_D)
        (\boldsymbol{\xi}\cdot\nabla\Omega_D)\hat{\bf R}
    }.
\end{align}

Given the eigenmodes $\{\boldsymbol{\xi}_j\}$ of a rigidly rotating planetary model, we make the ansatz that normal modes of the differentially rotating model can be expressed as a series expansion
$\boldsymbol{\xi}=\sum_jc_j\boldsymbol{\xi}_j$, with coefficients $c_j$ to be determined. Inserting this expansion, we write the momentum equation as
\begin{equation}
    \sum_jc_j\left(
        \mathcal{H}^R[\omega,\boldsymbol{\xi}_j]
        -\mathcal{H}^R[\omega_j,\boldsymbol{\xi}_j]
        +\mathcal{H}^D[\omega,\boldsymbol{\xi}_j]
    \right)
    =0,
\end{equation}
where superscripts $R$ and $D$ distinguish operators associated with rigid and differential rotation: 
\begin{align}
    \mathcal{H}^R[\omega,\boldsymbol{\xi}_j]
    &=\omega^2\mathcal{T}[\boldsymbol{\xi}_j]
    +2\omega \mathcal{W}^R[\boldsymbol{\xi}_j]
    +\mathcal{V}[\boldsymbol{\xi}_j],
\\
    \mathcal{H}^D[\omega,\boldsymbol{\xi}_j]
    &=2\omega \mathcal{W}^D[\boldsymbol{\xi}_j]
    +\mathcal{U}[\boldsymbol{\xi}_j],
\\
    \mathcal{W}^R[\boldsymbol{\xi}_j]
    &=\text{i}{\bf \Omega_S}\times \boldsymbol{\xi}_j,
\\
    \mathcal{W}^D[\boldsymbol{\xi}_j]
    &=\text{i}{\bf \Omega_D}\times \boldsymbol{\xi}_j
    -m\Omega_D\boldsymbol{\xi}_j,
\end{align}
such that $\mathcal{H}^R[\omega_j,\boldsymbol{\xi}_j]=0$, $\forall j$ (by construction), but $\mathcal{H}^R[\omega,\boldsymbol{\xi}_j]$ is not necessarily zero. Taking the inner product with an arbitrary $\boldsymbol{\xi}_i^*$ then produces a matrix equation ${\bf A\cdot c}=0$ for coefficient vectors $\bf c$. Writing 
$\langle\boldsymbol{\xi}_i,
\mathcal{L}[\boldsymbol{\xi}_j]\rangle
:=\mathcal{L}_{ij}$ for a given operator $\mathcal{L}$, $\bf A$ has matrix elements
\begin{equation}\label{eq:Ael}
    A_{ij}=(\omega^2-\omega_j^2)\mathcal{T}_{ij}
    +2(\omega-\omega_j)\mathcal{W}_{ij}^R
    +2\omega\mathcal{W}_{ij}^D
    +\mathcal{U}_{ij}.
\end{equation}
Note that $\mathcal{V}_{ij}$ cancels because it does not involve the (in general unequal) eigenfrequencies $\omega_j$ and $\omega$ of the rigidly and differentially rotating models, and because we have ignored modification of the background pressure, density and gravitational field by the differential rotation.

If the effects of differential rotation are small enough that the off-diagonal elements of ${\bf A}$ are negligible, then the equation ${\bf A\cdot c}=\sum_jA_{ij}c_j=0$ reduces to $A_{ii}c_i=0,$ from which predicted frequency shifts $\Delta\omega_i=\omega-\omega_i$ can be calculated for a given $\boldsymbol{\xi}_i$ as
\begin{align}\label{eq:drot_fshift}
    \Delta\omega_i
    &\simeq\frac{-(
        2\omega_i\mathcal{W}_{ii}^D
        +\mathcal{U}_{ii}^D
    )}{
    2\left(
        \omega_i\mathcal{T}_{ii}
        +\mathcal{W}_{ii}^R
        +\mathcal{W}_{ii}^D
    \right)
    }.
\end{align}
Such shifts only remain accurate in the absence of strong mode mixing. In general, we write ${\bf A}=\omega^2{\bf B}+\omega{\bf C}+{\bf D}$
and solve the quadratic generalized eigenvalue problem
\begin{equation}\label{eq:drot_mat}
    \left[\begin{matrix} 
        0       & {\bf D}  \\
        {\bf D} & {\bf C}
    \end{matrix}\right]\cdot
    \left[\begin{matrix}  
        {\bf c} \\ \omega {\bf c}
    \end{matrix}\right]
    =\omega\left[\begin{matrix} 
        {\bf D} & 0  \\
        0       & -{\bf B}
    \end{matrix}\right]\cdot
    \left[\begin{matrix}  
        {\bf c} \\ \omega {\bf c}
    \end{matrix}\right].
\end{equation}
Once calculated, the coefficient vectors ${\bf c}$ can be used to reconstruct approximations to the eigenfunctions of the differentially rotating system from the eigenmodes of the rigidly rotating system. 

We stress that this treatment of differential rotation is only approximate, not least because it ignores baroclinic modification of the equilibrium state. Our perturbative treatment also relies on the assumption that the oscillations of the differentially rotating planet can be represented by a truncated expansion in the modes of a rigidly rotating model. However, given that Saturn's zonal winds constitute such a small perturbation to the planet's bulk rotation, in this preliminary work we operate under the assumption that they only facilitate the interaction of different oscillations, without drastically altering the rigidly rotating mode spectrum.

\section{Results with rigid rotation}\label{sec:rig}
In this section we provide a qualitative description of the oscillation modes calculated non-perturbatively for our rigidly rotating models of Saturn, before moving on to a quantitative discussion of their modification by differential rotation in \autoref{sec:diff}.

\subsection{Mode characterization}
We focus on fundamental modes, and gravito-inertial modes with super-inertial frequencies $|\omega|>2\Omega_S.$ Acoustic ``p modes'' may have been detected in Saturn's gravity field \citep{Markham2020}, but possess frequencies large enough that the associated resonances lie interior to Saturn's rings. On the other hand, although the tidal excitation and damping of inertial waves with sub-inertial frequencies $|\omega|<2\Omega_S$ may be relevant to the rapid migration of Saturn's moons \citep{Fuller2016,Lainey2020}, such low-frequency gravito-inertial modes would produce resonances exterior to the C Ring. 

\begin{figure*}
    \centering
    \includegraphics[width=\textwidth]{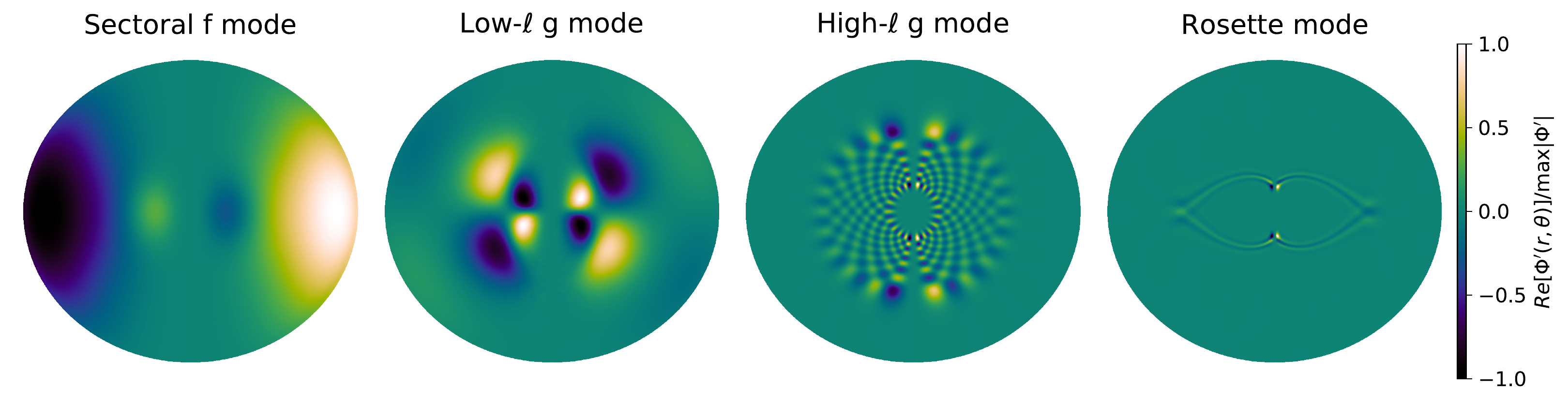}
    \caption{Meridional slices (cross-sections along the rotational axis) illustrating the gravitational perturbations for typical oscillations with azimuthal wavenumber $m=3$, for which $\Phi'(r,\theta,\phi=0)=-\Phi'(r,\theta,\phi=\pi)$. These modes were calculated for our model with $r_\text{stab}=0.70R_s$, but generally appear in all of our calculations (with slightly different frequencies). Many f modes and g modes retain eigenfunctions dominated by a single spherical harmonic (dominant $\ell=3,4,21$ for the left-hand panels), but rotation also produces rosette modes with more exotic angular structures (the right-most panel shows an example).}
    \label{fig:dPhdemo}
\end{figure*}

The color-plots in \autoref{fig:dPhdemo} show the meridional $(r,\theta)$ structure of the gravitational perturbation associated with a selection of oscillation modes calculated for our model with stable stratification extending to $r_\text{stab}=0.70R_S$ at the equator. The left-most color-plot illustrates the eigenfunction of the $m=3$ sectoral f mode. In our models, the f modes with $\ell\lesssim4$ reside in both the convective envelope and the stably stratified interior, where they take on a gravito-inertial character. Their wavefunctions strongly overlap with those of the lowest-order g modes (independent of frequency degeneracy or rotational mixing), to the point of eliminating any meaningful physical distinction; we simply identify the oscillations with the largest surface gravitational perturbations under \autoref{eq:norm} as the f modes. The f modes' confinement to the envelope becomes stricter at higher spherical harmonic degrees, resolving their distinction from the g modes and making the oscillations more sensitive to Saturn's zonal winds (see \autoref{sec:diff}).

\begin{figure*}
    \centering
    \includegraphics[width=\textwidth]{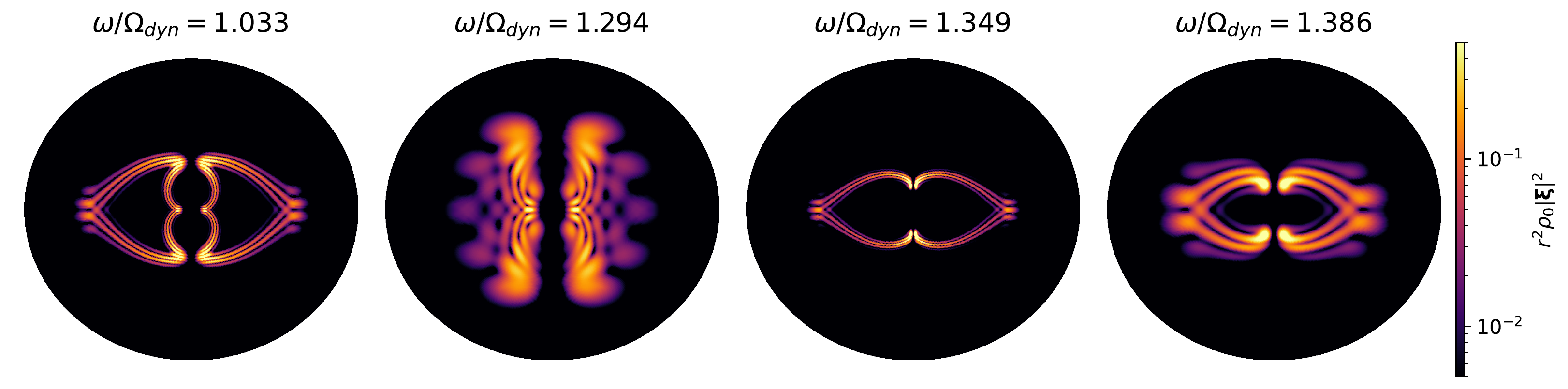}
    \caption{Meridional slices illustrating kinetic energy distributions (arbitrarily normalized and multiplied by $r^2$) for $m=3$ rosette modes calculated from our fiducial model with $r_\text{stab}=0.70R_s$. The inner (outer) panels show rosette modes with even (odd) equatorial parity. In our models, the two modes shown in the middle panels generically appear as bookends to a sequence of modes with densely spaced frequencies that transition from vertical to horizontal alignment.}
    \label{fig:rosetteKE}
\end{figure*}

\begin{figure*}
    \centering
    \includegraphics[width=\textwidth]{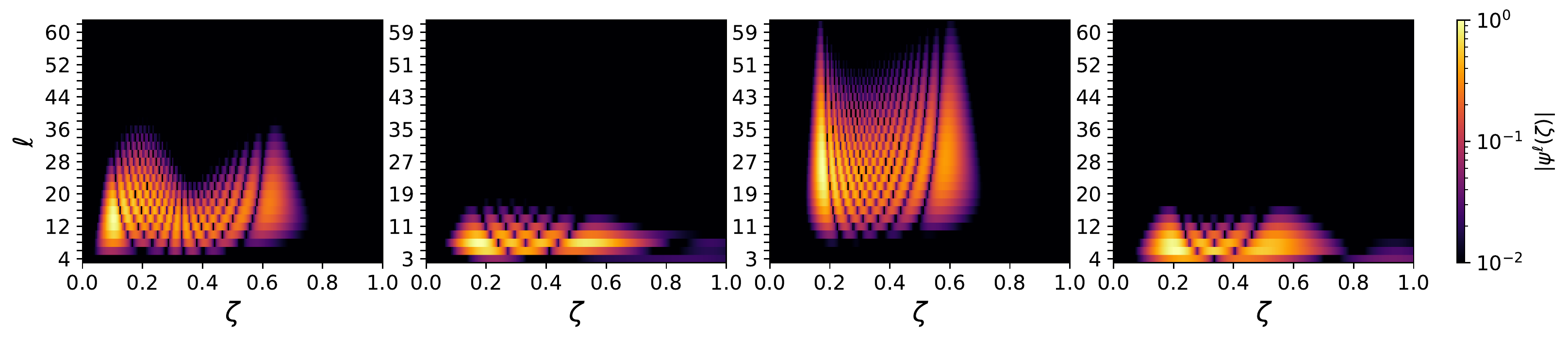}
    \caption{Color-plots illustrating the spectral composition of the $m=3$ rosette modes shown in \autoref{fig:rosetteKE}. The heatmaps show, at each spherical harmonic degree $\ell$ (y-axis) and quasi-radial coordinate $\zeta$ (x-axis), the (arbitrarily normalized) amplitude of the spectral coefficients $\psi^\ell(\zeta)$ in an expansion of the modes' gravitational perturbations in spherical harmonics (see \autoref{app:npert} for variable definitions; for a given mode, all the perturbed fluid variables share similar spectral compositions). The panels demonstrate that the eigenfunctions of these oscillations involve many spherical harmonic degrees. Note that the equatorial symmetry of purely rigid rotation enforces strictly odd (even) equatorial parity for the modes with spectral compositions shown in the outer (inner) panels, so the y-axes include $\ell=4,6,8,...$ ($\ell=3,5,7,...$).}
    \label{fig:rosetteSpec}
\end{figure*}

Although the stably stratified interiors of our planetary models are simple, Saturn's rapid rotation complicates the g mode spectrum. As found by \citet{Ballot2010,Ballot2012}, in the super-inertial frequency range of interest our calculations produce both conventional g modes (\autoref{fig:dPhdemo}, middle left and middle right), and more exotic rosette modes (\autoref{fig:dPhdemo}, right; \autoref{fig:rosetteKE}; \autoref{fig:rosetteSpec}). The former oscillations retain a well-defined number of nodes in the quasi-radial and polar directions, and can be clearly identified with non-rotating counterparts characterized by a single spherical harmonic (thoughout, we use $\ell\sim$ to denote this correspondence). For these modes, rotation acts primarily to concentrate the oscillations toward the equator \citep{Townsend2003}. 

In contrast, rosette modes take on an altogether different angular structure, characterized by rosette patterns traced in the distributions of kinetic energy (\autoref{fig:rosetteKE} shows examples for a variety of $m=3$ rosettes) that inherently involve a coupling across many spherical harmonics (demonstrated by \autoref{fig:rosetteSpec}). Non-perturbative, or at least degenerate perturbative treatments of the partial differential equations are therefore required to calculate these oscillations. Rosette modes can be identified with the rotational mixing of sequences of g modes that in the non-rotating regime have nearly degenerate frequencies, successive spherical harmonic degrees $\ell$ separated by $2$, and different radial quantum numbers $n$ \citep[see Fig. 1 in ][]{Takata2013}. They also appear as stable periodic orbits in ray-dynamic calculations \citep{Prat2016,Prat2018}. 

While the appearance and properties of families of rosette modes depend non-trivially on near-degeneracies in the non-rotating g mode spectrum, asymptotic analyses do predict regularities \citep{Takata2014}, some of which we observe in our calculations. For example, the $m=3$ modes with kinetic energies illustrated by the two middle panels of \autoref{fig:rosetteKE} always appear at either end of a sequence with rosette patterns that transition from vertical to horizontal alignment as (closely spaced) frequencies increase. This corresponds to an increase in the real parameter ``q'' considered by \citet{Takata2014} in their JWKB analysis (see their Fig. 2). 

For the purposes of our investigation the rosette modes play an effectively similar role to more conventional high-degree g modes \citep[indeed, the distinction becomes blurry for prograde, non-axisymmetric modes; ][]{Saio2014}. We note, though, that those with vertically rather than horizontally aligned rosette patterns generally couple more strongly with the sectoral f modes, likely due to spectral compositions involving lower-degree spherical harmonics (compare the two middle panels in \autoref{fig:rosetteSpec}).

\begin{figure}
    \centering
    \includegraphics[width=\columnwidth]{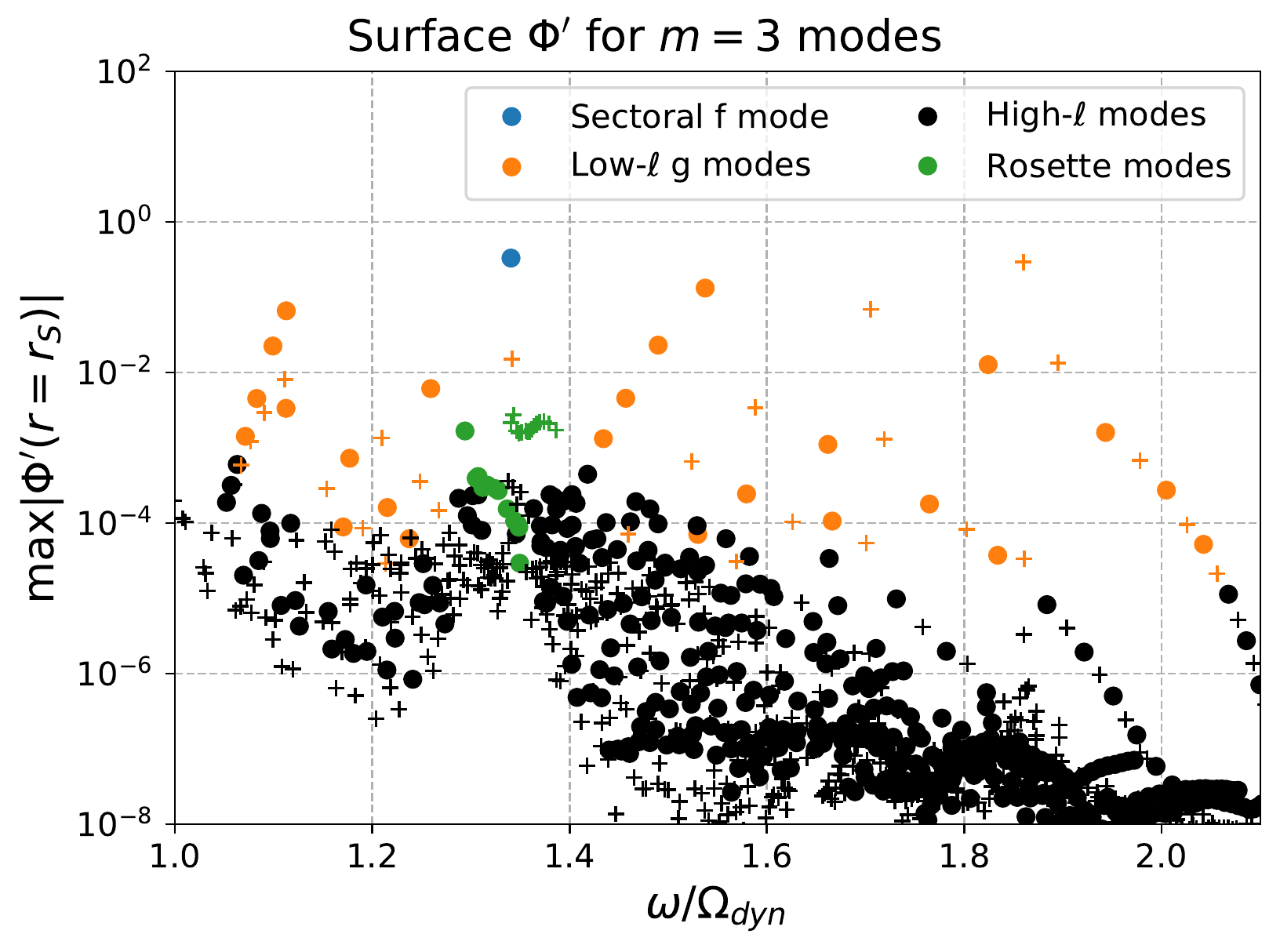}
    \caption{Maximum surface gravitational perturbations plotted against frequencies for identically normalized, $m=3$ modes calculated from our rigidly rotating model with $r_\text{stab}=0.70R_s$. Circles (plus signs) denote modes with even (odd) equatorial parity (even/odd $\ell-m$). Under the normalization of \autoref{eq:norm}, the sectoral f mode (blue point; see also \autoref{fig:dPhdemo}, left) has the largest surface $\Phi',$ closely followed by g modes with eigenfunctions dominated by relatively low spherical harmonic degrees $\ell$ and radial orders $n$ (orange points; \autoref{fig:dPhdemo}, middle). Rosette modes (green points; \autoref{fig:dPhdemo}, right and \autoref{fig:rosetteKE}) have smaller surface perturbations, but manifest as sequences with densely spaced frequencies (in this case close to that of the sectoral f mode).}
    \label{fig:dPh_v_om}
\end{figure}

\subsection{Surface gravitational perturbations}
\autoref{fig:dPh_v_om} plots the maximum surface gravitational perturbations for $m=3$ oscillations with both even (circles) and odd (plus signs) equatorial parity, calculated from the same fiducial model (with $r_\text{stab}=0.70R_s$) as \autoref{fig:dPhdemo}. The blue point denotes the sectoral $m=3$ f mode, which has the largest surface $\Phi'$ under the normalization of \autoref{eq:norm}. The orange points denote relatively low-degree, low-radial order ($\ell,n\lesssim12$), conventional g modes like that shown in \autoref{fig:dPhdemo} (middle left); their maximum surface values of $\Phi'$ come close to that of the f mode. On the other hand, rosette modes (two example sequences of which are indicated by green points) and other high-$\ell$ oscillations (black points) carry the majority of their kinetic energy in the stably stratified trapping cavity (i.e., they possess higher mode inertias). Their surface gravitational perturbations are consequently smaller relative to mode energies.

\citet{Fuller2014b} showed that avoided crossings between f modes and g modes with nearly equal frequencies can enhance the surface amplitudes of the latter. This is particularly true for the conventional g modes with relatively low $\ell$ and $n$, which in our calculations already possess large surface $\Phi'$ \citep[our models have much wider trapping cavities than those of ][so the g modes face a smaller distance over which to evanesce in the envelope]{Fuller2014b}. However, as demonstrated by the orange points in \autoref{fig:dPh_v_om}, these low-degree g modes are sparsely distributed in frequency. Additionally, the strength of frequency ``repulsion'' between the sectoral f mode and low-degree g modes near avoided crossings makes a close frequency splitting between an f mode and multiple low-degree g modes of the same equatorial parity unlikely.

Rotational mixing with high-$\ell$ g modes presents a possible alternative explanation for the observed triplet of $m=3$ density waves, since (as also demonstrated by the green and black points in \autoref{fig:dPh_v_om}) the high-degree g modes are very densely packed in frequency space; the rosette modes, in particular, appear in sequences with frequencies separated by a few degrees per day (comparable to the finest frequency splitting observed for $m=3$). For models similar to our fiducial case with $r_\text{stab}=0.70R_S$, the two rosette sequences shown by green points in \autoref{fig:dPh_v_om} also appear suggestively close in frequency to the sectoral $m=3$ f mode (replacing the conventional high-$\ell$ g modes that mix together to form them). The significance of this proximity should not be over-interpreted, however; rosette modes appear elsewhere in the dense spectrum of high-$\ell$ modes shown in \autoref{fig:dPh_v_om}, and the two highlighted sequences shift in frequency with changes to the background stratification of the model.

Further, as noted by \citet{Fuller2014b} rotational mixing between the f modes and high-degree g modes is weak. Although \autoref{fig:dPh_v_om} indicates an enhancement for oscillations with frequencies close to $\omega\sim1.35\Omega_\text{dyn}$ that is common to our models, the surface gravitational perturbations of high-$\ell$, $m=3$ g modes and rosette modes generally fall several orders of magnitude below that of the sectoral f mode for all our calculations with purely rigid rotation (at least under the normalization of \autoref{eq:norm}).

\begin{figure*}
    \centering
    \includegraphics[width=\columnwidth]{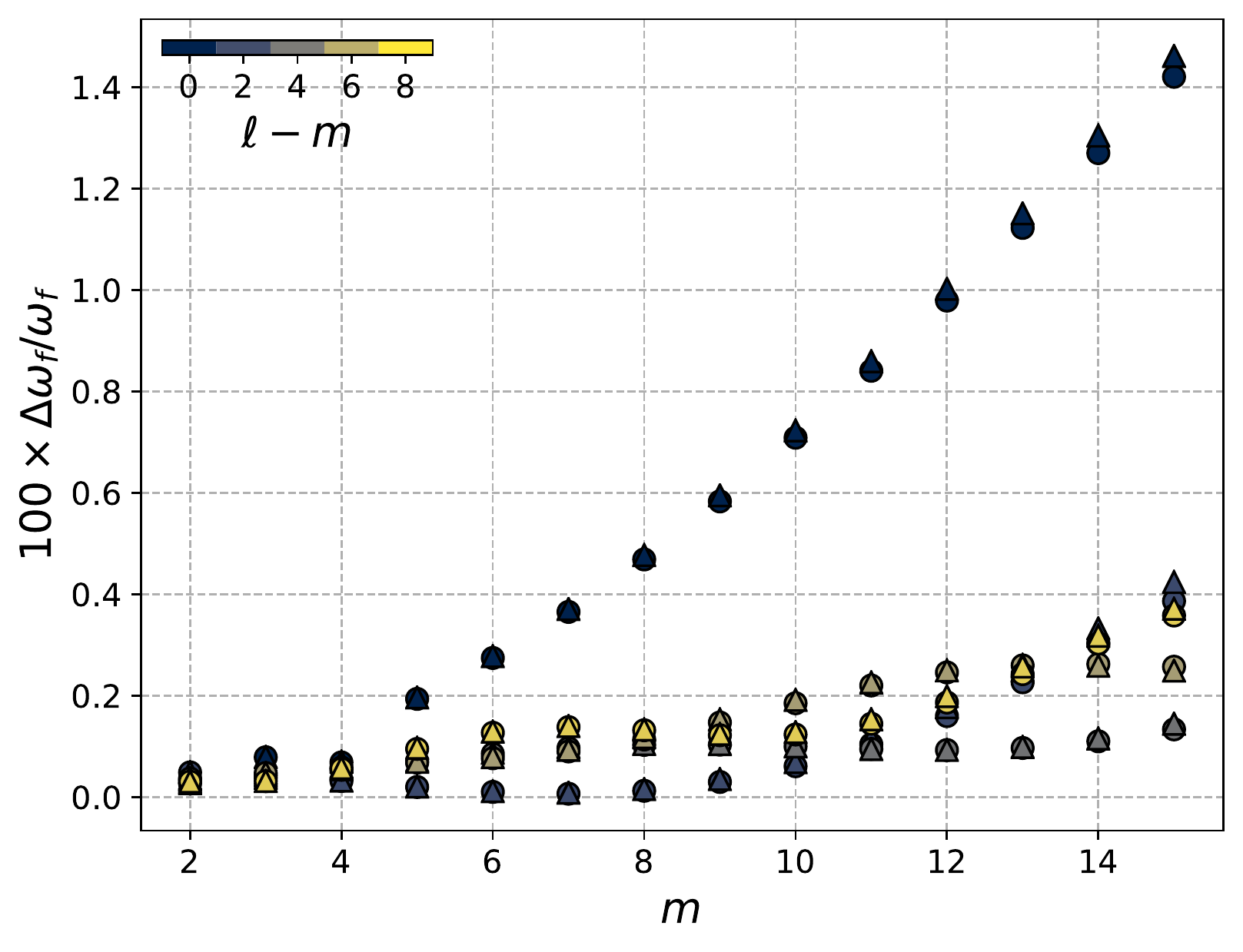}
    \includegraphics[width=\columnwidth]{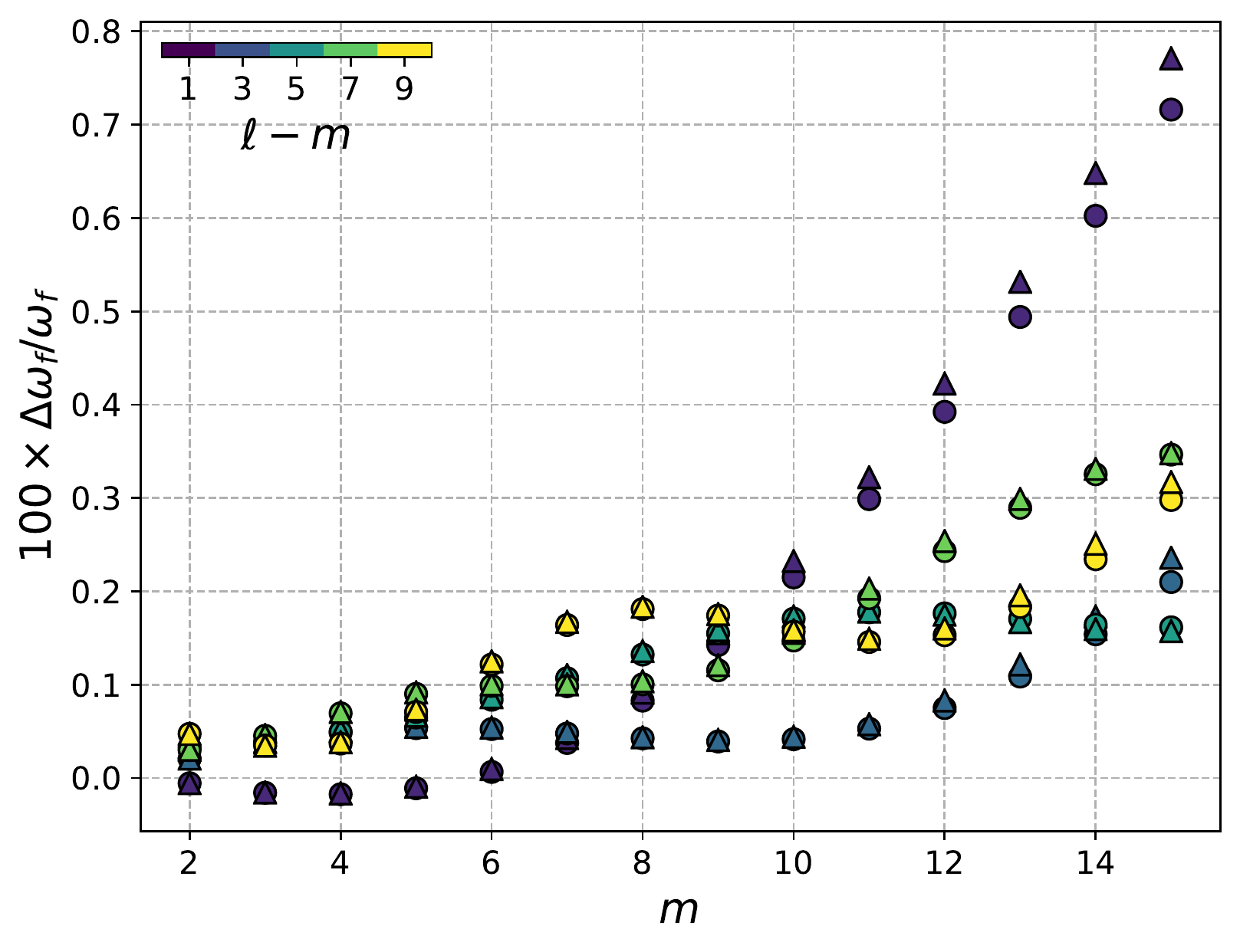}
    \caption{Per cent changes in frequency due to differential rotation for f modes with predominantly even (left) and odd (right) equatorial parity, calculated from our fiducial ($r_\text{stab}=0.70R_s$) model assuming a zonal wind decay depth of $d=0.125R_S\simeq7.5\times 10^3\text{km}$. The circles (triangles) denote frequencies calculated numerically by solving \autoref{eq:drot_mat} (from the approximate \autoref{eq:drot_fshift}).}
    \label{fig:rst68_fmode_freqshift}
\end{figure*}
\begin{figure*}
    \centering
    \includegraphics[width=\textwidth]{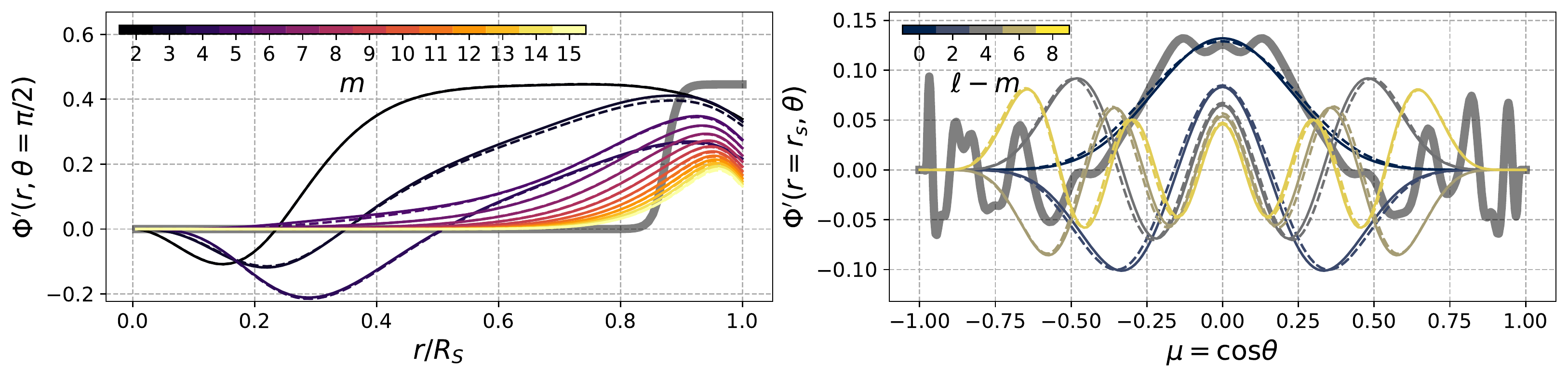}
    \caption{    
    Left: equatorial profiles of $\Phi'$ for the sectoral ($\ell- m\sim0$) f modes, calculated from our $r_\text{stab}=0.70R_s$ model with/without the effects of differential rotation (dashed/solid lines), and superimposed on the (scaled) decay profile assumed for the zonal winds (thick gray line). Right: surface profiles of $\Phi'$ for $m=15$ f modes with predominantly even equatorial parity (even $\ell-m$), plotted over the (scaled) surface differential rotation profile $\Omega_D$. The left-hand panel demonstrates increasing confinement toward the envelope, likely responsible for larger frequency shifts at larger $\ell\sim m$. Meanwhile, the right-hand panel illustrates the overlap between the confinement of the sectoral modes near $\mu=0$ and Saturn's equatorial jet, which explains the sectoral modes' larger increases in frequency (see \autoref{fig:rst68_fmode_freqshift}, left)
    .}\label{fig:rst68_fmode_eigfuns}
\end{figure*}

\begin{figure}
    \centering
     \includegraphics[width=\columnwidth]{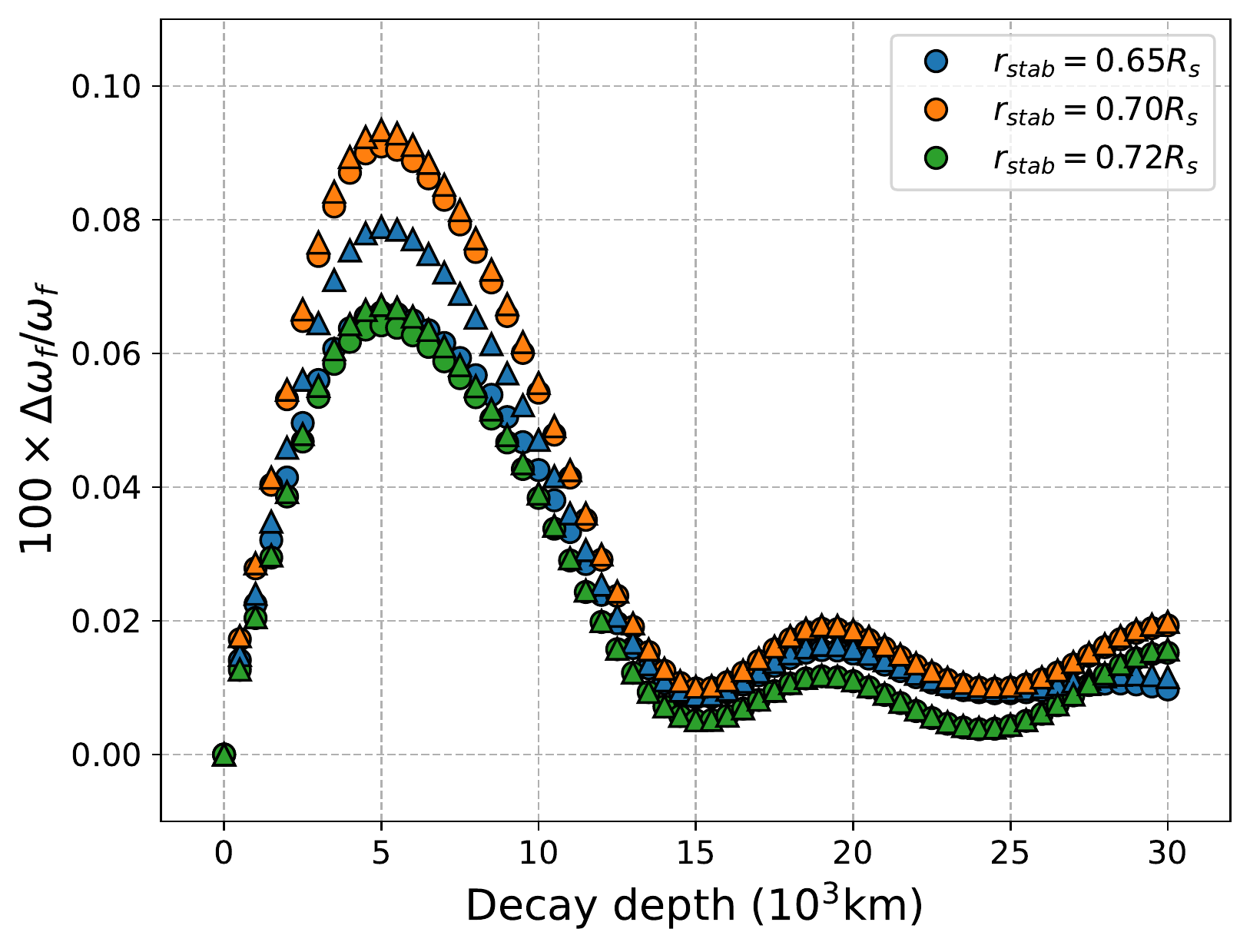}
    \caption{Per cent frequency changes for the sectoral $\ell\sim m=3$ f mode as a function of the wind decay depth, calculated (with equatorially symmetric wind profiles) for the three models considered in this paper. Triangles denote predictions from \autoref{eq:drot_fshift}, which can lose accuracy when modes are strongly mixed (this occurs for $r_\text{stab}=0.65R_S$ near $d\sim5\times10^3$km).
    The non-monotonic variation occurs as Saturn's high-latitude winds descend deeply enough to interact with the oscillation.}
    \label{fig:m3fmode_depthvar}
\end{figure}

\section{Results with differential rotation}\label{sec:diff}
Saturn's zonal winds are strong by terrestrial standards ($\lesssim 300\text{m}/\text{s}$), but modest in comparison to both the planet's bulk rotation rate and its dynamical frequency ($\Omega_D\lesssim0.03\Omega_S\sim0.01\Omega_\text{dyn}$). As a result, we find that the associated differential rotation modifies Saturn's oscillations subtly. 

\subsection{Frequency shifts}
The zonal winds directly affect f modes more strongly than g modes. \autoref{fig:rst68_fmode_freqshift} plots the per cent changes in frequency produced by a zonal wind with a decay depth of $d=0.125R_S\simeq7.5\times10^3\text{km}$, for f modes with even (left) and odd (right) equatorial parity calculated from our fiducial model with $r_\text{stab}=0.70R_s$. The points and plus signs respectively denote frequency shifts calculated by solving \autoref{eq:drot_mat}, and from the approximate frequency shift formula given by \autoref{eq:drot_fshift}. Shifts in frequency generally increase with azimuthal wavenumber, a trend explained by the increasingly strict confinement of eigenfunctions to the envelope with increases in the dominant spherical harmonic degree. The left-hand panel in \autoref{fig:rst68_fmode_eigfuns} shows the equatorial profiles of $\Phi'(r,\theta=\pi/2)$ for the sectoral f modes, illustrating this stricter confinement for larger $\ell\sim m$. The faint gray line plots the decay profile enforced for the winds (scaled to match the amplitude of the eigenfunctions), demonstrating the domain of influence for the differential rotation profile.

\autoref{fig:rst68_fmode_eigfuns} (right) shows surface profiles of  $\Phi'(r=r_s,\theta)$ for $m=15$ f modes, demonstrating an additional geometric consideration: Saturn's rapid rotation enhances \emph{equatorial} confinement, especially for the sectoral f modes. As a result, the high-$\ell\sim m$ sectoral f modes' localization coincides almost exactly with Saturn's rapid equatorial jet [the scaled surface profile of $\Omega_D(r=r_s)$ is plotted in gray], while the eigenfunctions of the higher-degree oscillations extend to latitudes dominated by alternating regions of under-rotation and super-rotation. These differences in equatorial confinement provide a natural explanation for the monotonic relationship between frequency shift and $m$ for the sectoral oscillations, vis-\'a-vis the non-monotonic variation for the other f modes. 

These geometrical considerations extend to the low-degree sectoral f modes, even though they are less affected by the zonal winds. \autoref{fig:m3fmode_depthvar} plots the per cent frequency changes for the $m=3$ sectoral f mode calculated for all of the models shown in \autoref{fig:models}, for decay depths increasing from zero to an exaggerated $d=30\times10^3$km. The calculations shown in this plot alone were performed with the antisymmetric component of the wind suppressed, in order to avoid discontinuities deep in the interior (see \autoref{sec:zwind}). The non-monotonic variation in frequencies with increasing depth can be directly associated with alternating regions of under-rotation and super-rotation at higher latitudes extending deeply enough to overlap with the oscillations' eigenfunctions. 

In particular, the relatively large decrease in the frequency shifts plotted in \autoref{fig:m3fmode_depthvar} (between $d\sim5\times10^3$km and $d\sim15\times10^3$km for all three models) is due to an increased overlap of the $m=3$ f mode eigenfunctions with under-rotation at $|\mu|\sim0.5-0.6$ on the surface (see \autoref{fig:satWind}, left) as it extends to meet the equator at $r(\mu=0)\sim0.8R_S$. This reduction of the frequency enhancement by zonal winds is relevant given the inference of a sub-corotating layer based on measurements of Saturn's gravitational moments \citep{Iess2019}, but does not significantly alter the frequency shifts of the high-degree sectoral modes shown in \autoref{fig:rst68_fmode_freqshift} (since their eigenfunctions are concentrated toward the surface).

\subsection{Mode mixing}
The zonal wind profiles used in our calculations have little-to-no \emph{direct} impact on most of the g modes of our models. For all but the lowest degrees $\ell$ and radial orders $n$, the frequencies of isolated g modes remain largely unaltered except for very deep decay depths $d\gtrsim20\times10^3\text{km}$. This is not surprising, since all but the lowest order oscillations are confined strictly to the stably stratified cavity in the deep interior, where we impose negligible differential rotation. However, high-degree g modes and rosette modes can be affected \emph{indirectly} through rotational mixing with f modes whose frequencies are altered by the zonal winds. In the following subsections we describe a few examples of such interactions, before discussing observational implications.

\begin{figure*}
    \centering
    \includegraphics[width=\textwidth]{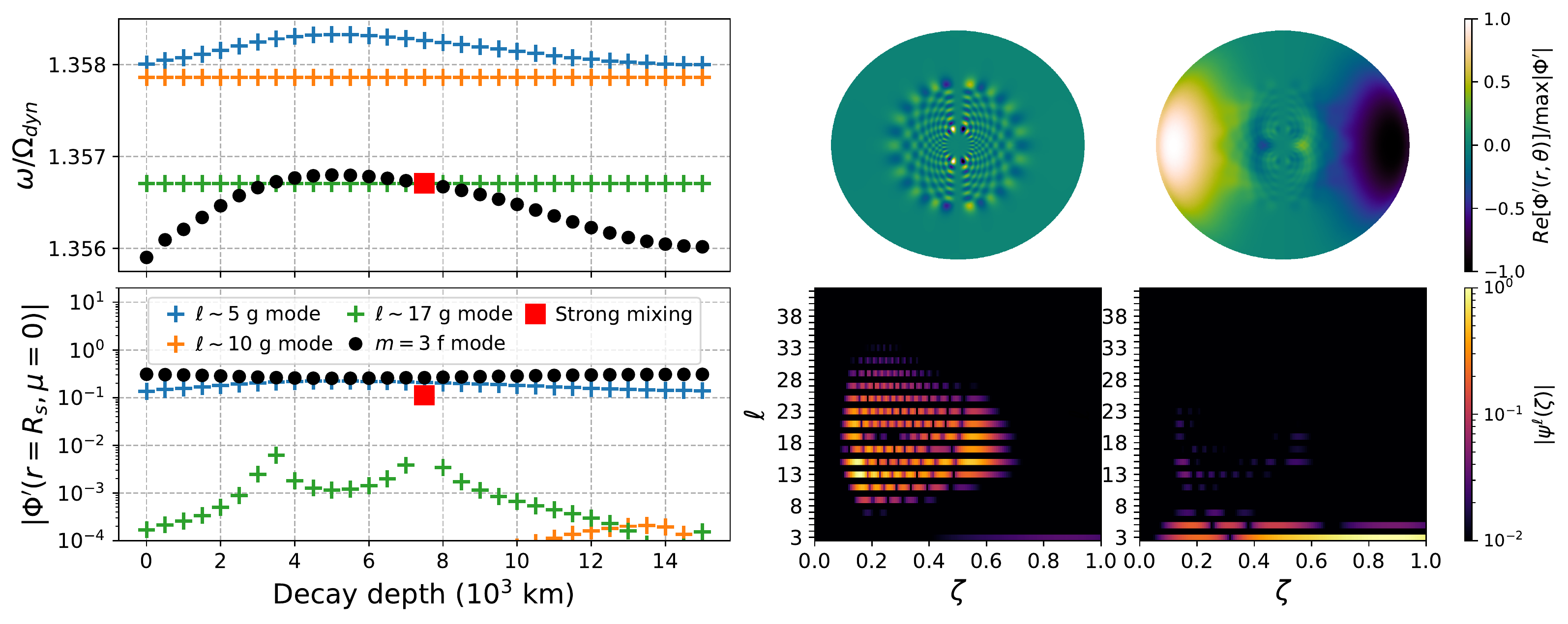}    
    \caption{Left: changes in frequency (top) and equatorial surface gravitational perturbation (bottom) with zonal wind decay depth for the $m=3$ f mode (black points) and a few nearby g modes (plus signs) calculated from our model with $r_\text{stab}=0.65R_S$. The zonal winds have little direct effect on the frequencies of the high-degree g modes, but they gain enhanced surface $\Phi'$ from avoided crossings encountered as the f mode's frequency changes with decay depth (e.g., the red square at $d=7.5\times10^3$km). Right: meridional slices (top) and spectral decompositions (bottom; note that unlike the decompositions shown in \autoref{fig:rosetteSpec}, these include both even and odd $\ell$) showing the gravitational perturbation of the $\ell\sim17$ mode denoted by the green plus signs and red square in the left-hand panels, both in the absence of differential rotation (left) and during the avoided crossing at $d=7.5\times10^3$km (right). These color-plots illustrate the enhancement of the $\ell=3$ component of the $\ell\sim17$ mode due to mixing with the f mode.}
    \label{fig:rst65_avcross}
\end{figure*}
\begin{figure*}
    \centering
    \includegraphics[width=0.75\textwidth]{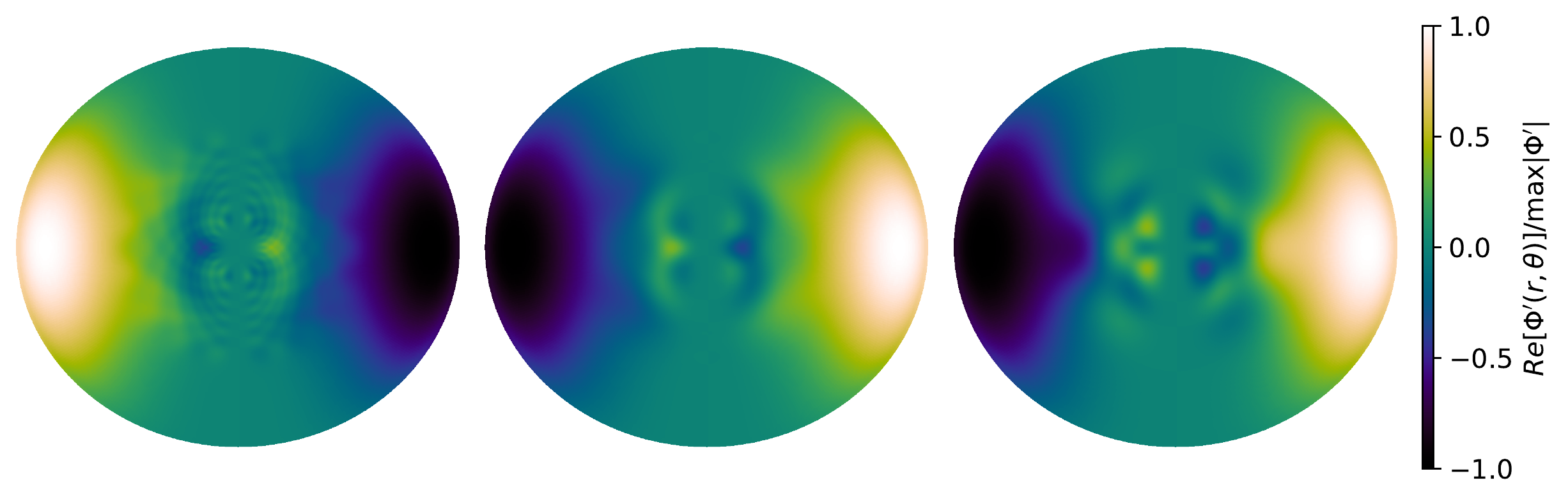}
    \caption{Meridional slices showing $\Phi'(r,\theta)$ for the three modes from \autoref{fig:rst65_avcross} with the largest surface $|\Phi'|$ at $d=7.5\times 10^3\text{km}$. In this three-mode interaction, the sectoral $m=3$ f mode (center) enhances the $\ell=3$ component of the gravitational perturbation for two g modes with $\ell\sim17$ (left) and $\ell\sim5$ (right).}
    \label{fig:rst65_triplet}
\end{figure*}

\begin{figure*}
    \centering
    \includegraphics[width=\textwidth]{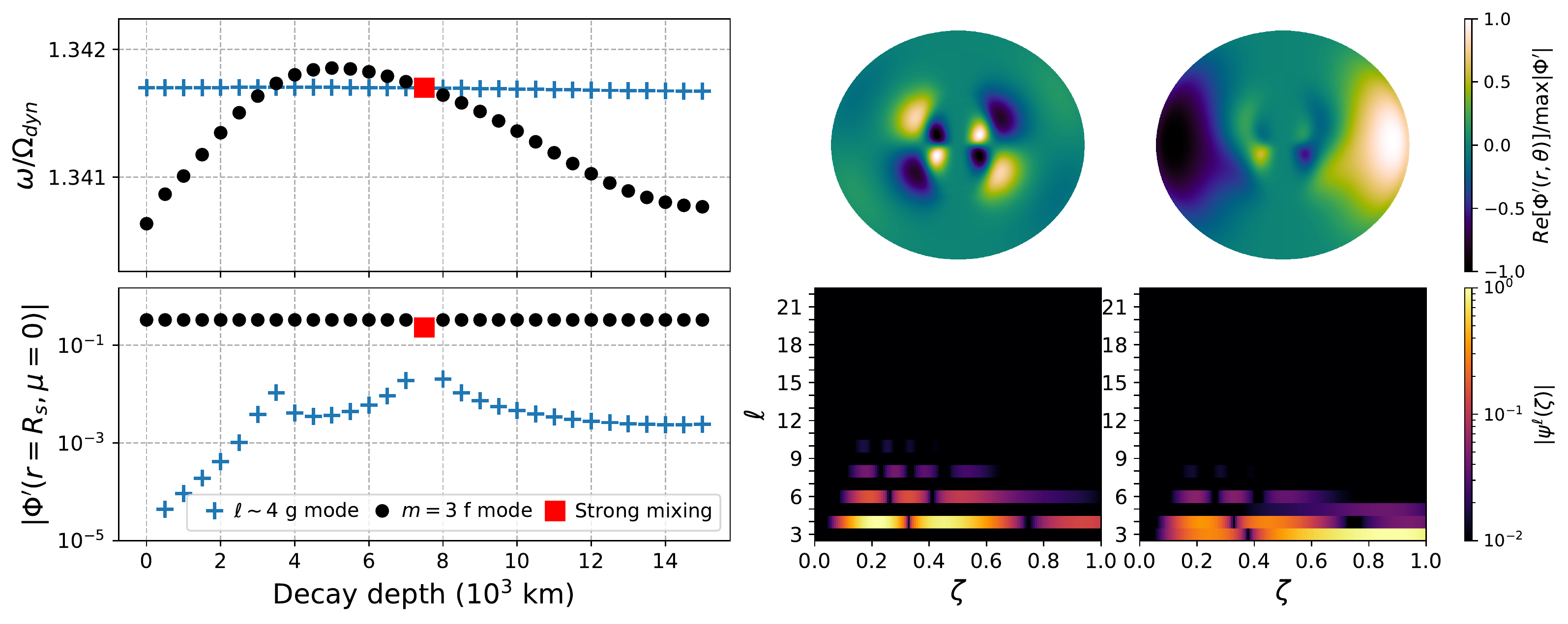}
    \caption{Same as the panels in \autoref{fig:rst65_avcross}, but for our fiducial model with $r_\text{stab}=0.70R_S,$ for which the frequency of the $m=3,$ $\ell\sim4$, $n=2$ g mode (with odd equatorial parity) falls close to that of the sectoral $m=3$ f mode. The antisymmetric part of Saturn's zonal winds is weak, but sufficient to mix this odd-parity g mode with the even-parity f mode, producing two oscillations with eigenfunctions involving both even and odd harmonics (as shown for the g mode in the right-hand spectral decomposition). Asymmetric oscillations like that shown in the top-right panel may have a unique observational signature (see \autoref{sec:simul}), namely the simultaneous excitation of both density and bending waves at different locations in the C Ring.}
    \label{fig:rst68_avcross}
\end{figure*}

\subsubsection{Mixing with high-degree g modes and rosettes}
\autoref{fig:rst65_avcross} illustrates avoided crossings encountered by the $m=3$ f mode as the zonal wind decay depth increases for our model with a Brunt-V\"ais\"al\"a profile extending to $r_\text{stab}=0.65R_S$ at the equator. The left-hand panels plot the frequencies (top) and maximum surface gravitational perturbations (bottom) for the $m=3$ sectoral f mode (black circles) and nearby g modes (blue, orange and green plus signs). Since the high-degree g modes and rosette modes are distributed densely in frequency-space, we find that for any given model, at least one decay depth $5\times10^3\text{km}\lesssim d\lesssim 10^4\text{km}$ produces an avoided crossing with the sectoral $m=3$ f mode. During these avoided crossings, the f mode and g mode frequencies nearly overlap, and the oscillations trade character. Close to an avoided crossing the f mode imparts a larger gravitational perturbation to the g mode.

This is illustrated by the color-plots in the right-hand panels of \autoref{fig:rst65_avcross}, which compare meridional (top) slices and spectral decompositions (bottom) of $\Phi'(r,\theta)$ both with (right) and without (left) differential rotation, for the relatively high-degree ($\ell\sim 17$) g mode with frequency and surface $\Phi'$ indicated by the red square. Note that unlike \autoref{fig:rosetteSpec}, the spectral decompositions shown in \autoref{fig:rst65_avcross} include even-$\ell$ components, which remain negligible for this $m=3$ mode with even equatorial parity (even $\ell-m$). Both the meridional slices and the spectral decompositions show an enhancement of the $\ell=3$ component of the mode's eigenfunction.

Such avoided crossings provide an attractive explanation for observed $m=3$ density waves with frequencies split by less than one per cent. A major caveat is that, as shown in \autoref{fig:rst65_avcross}, close matches in frequencies (with differences $\lesssim0.01-0.1$ per cent) are still required to endow the high-$\ell$ g modes and most rosette modes with surface gravitational perturbations comparable to that of the $m=3$ f mode (under the normalization of \autoref{eq:norm}, at least). Consequently, while our calculations for rigidly rotating models typically possess a spectrum of high-$\ell$ g modes and rosette modes that is dense enough near the f mode's frequency to produce at least one avoided crossing at a zonal wind decay depth $\lesssim 10^4\text{km}$, explaining the observed $m=3$ triplet likely requires at least one other flavor of mode mixing. 

For the model with $r_\text{stab}=0.65R_S$ considered in \autoref{fig:rst65_avcross}, the $m=3$ f mode additionally falls close in frequency to a lower-order g mode, which in the non-rotating regime can be identified with spherical harmonic degree $\ell=5$ and radial order $n=3$. Because this mode's solution comprises lower-$\ell$ spherical harmonics, the oscillation does not need to be as close in frequency to the f mode to gain a comparable surface amplitude. In fact, \autoref{fig:rst65_avcross} (top left) demonstrates clear frequency repulsion between this lower-degree g mode and the sectoral f mode, which prohibits the two oscillations from coming closer than $\sim0.1$ per cent in frequency separation. \autoref{fig:m3fmode_depthvar} also illustrates this repulsion, which causes a deviation between frequency shifts calculated in full (blue circles) and from the approximate \autoref{eq:drot_fshift} (blue triangles) near $d\sim5\times10^3$km. 

The meridional slices in \autoref{fig:rst65_triplet} illustrate the eigenfunctions for the three modes from \autoref{fig:rst65_avcross} with the largest surface $\Phi'$ at $d=7.5\times10^3$km. This decay depth happens to agree with current estimates \citep{Galanti2021}, but in our calculations $d$ is primarily significant for determining the modes' proximity in frequency space; the enhancement of mixing between f modes and g modes by differential rotation does not seem to depend heavily on decay depths between $5\times10^3-10^4$km.

As we discuss in \autoref{sec:dtau}, this type of interaction between i) the $m=3$ sectoral f mode, ii) a relatively low-degree g mode mode, and iii) a higher-degree g mode or rosette mode could conceivably explain the observed $m=3$ triplet of density waves. However, in general we find that engineering this type of three-mode interaction still requires fine-tuning of models (in order to place a lower-$\ell$ g mode close in frequency to the f mode), which in this case worsens agreement with the observed $m=2$ density waves (see \autoref{sec:disc}).

\subsubsection{Equatorial parity-mixing}\label{sec:pmix}
\autoref{fig:rst68_avcross} illustrates another very interesting possible explanation for the three finely split $m=3$ density waves. The figure shows similar calculations to those in \autoref{fig:rst65_avcross}, but for our fiducial model with a profile for the buoyancy frequency that extends to $r_\text{stab}=0.70R_s$. For this model (and many others like it), the $m=3$ f mode falls very close in frequency to the $\ell\sim 4$, $m=3,$ $n=2$ g mode (see the orange plus sign with the largest surface $\Phi'$ near $\omega\sim1.35\Omega_\text{dyn}$ in \autoref{fig:dPh_v_om}). We find that the antisymmetric part of a differential rotation profile like that shown in \autoref{fig:satWind} provides a weak coupling between these two oscillations with differing equatorial parity. The result is an avoided crossing between modes with \emph{asymmetric} equatorial structure, their eigenfunctions composed of both even and odd-degree spherical harmonics. 

This coupling is weaker than that between f modes and g modes with the same equatorial parity, and unlikely to cause significant mixing between a sectoral f mode and an odd-parity g mode of high-degree. Our model with $r_\text{stab}=0.72R_S$ does exhibit parity-mixing between the sectoral $m=3$ f mode and an odd-parity, $\ell\sim18$ g mode (along with a simultaneous coupling with an even-parity, $\ell\sim11$ g mode). However, the eigenfunction of this higher-degree, odd-parity g mode only gains a significant $\ell=3$ component when its frequency differs from that of the f mode by $\lesssim 10^{-4}$ per cent. Aside from requiring a significant amount of fine-tuning, such a small frequency separation so enhances the concentration of kinetic energy in the interior for the f mode (i.e., increases its inertia) that it no longer has the largest surface $\Phi'$ under the normalization of \autoref{eq:norm}.

Low-degree, odd-parity g modes, on the other hand, do not suffer the same frequency repulsion from the sectoral f modes as sectoral g modes. At the same time, they are not as disadvantaged by a high concentration of kinetic energy in the interior as high-degree g modes and rosette modes. Therefore, while substantial mixing between a sectoral f mode and a high-degree, odd-parity g mode is unlikely, a three-mode interaction involving i) the $m=3$ f mode, ii) an odd-parity, \emph{low-degree} g mode, and iii) an even-parity g mode may provide a viable explanation for the finely split triplet. We discuss the observational implications of such an interaction, and how it might be confirmed, in \autoref{sec:dtau}.

\section{Discussion}\label{sec:disc}
In this section we discuss the implications of the results described in \autoref{sec:rig} and \autoref{sec:diff}, and their relation to the waves observed in Saturn's C Ring.

\subsection{Comparison with observed frequencies}\label{sec:obs}
\begin{figure*}
    \centering
    \includegraphics[width=\columnwidth]{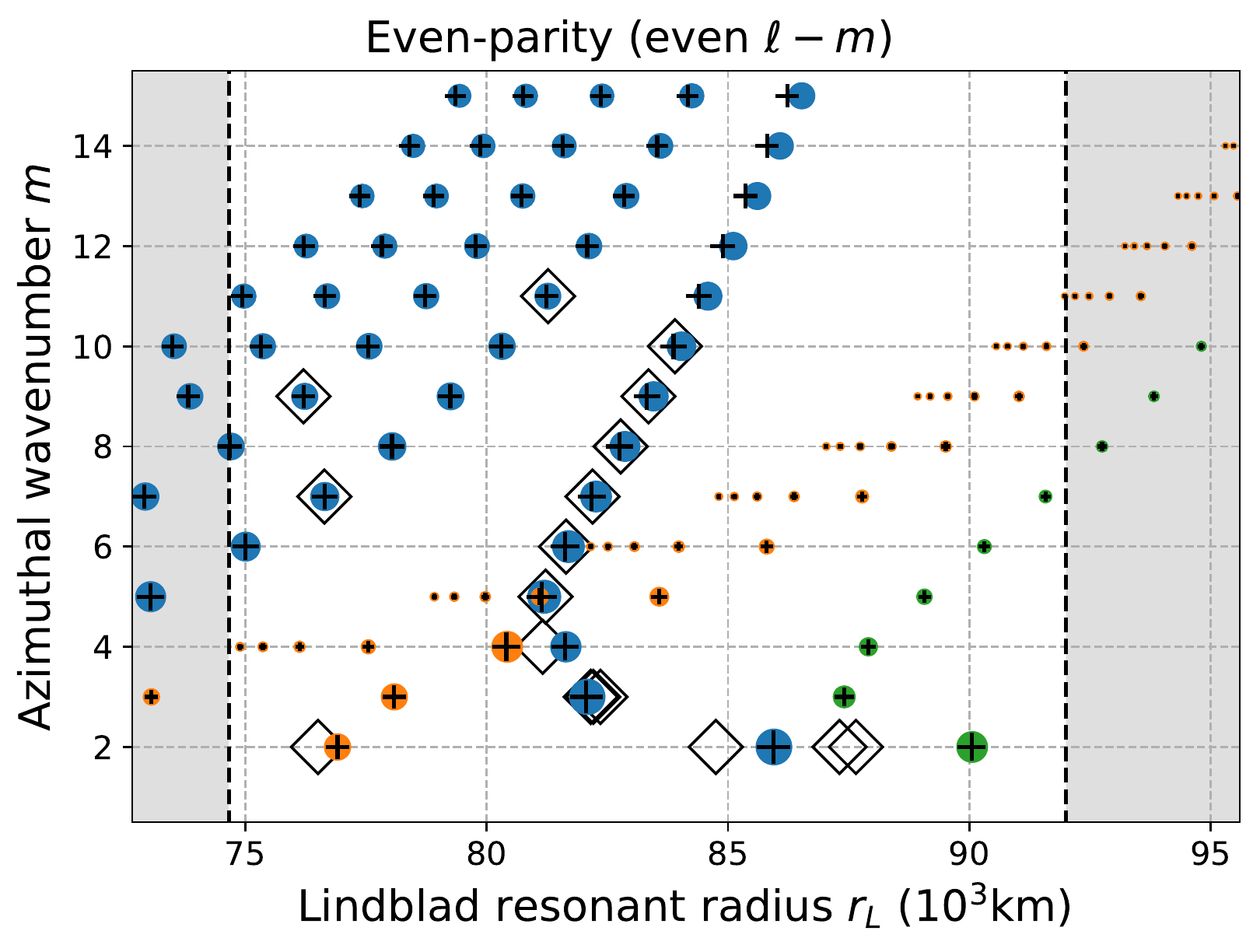}
    \includegraphics[width=\columnwidth]{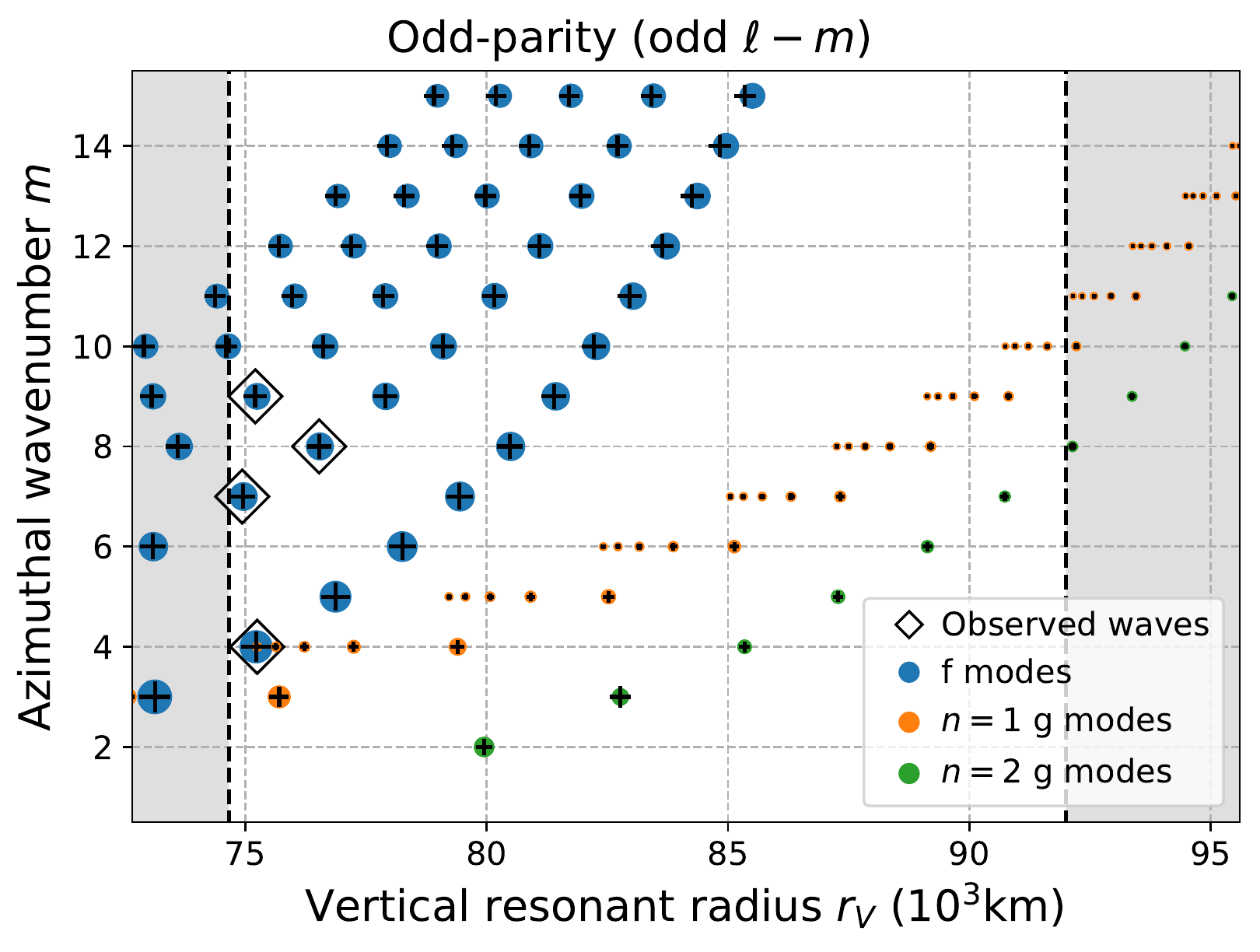}
    \caption{Left: azimuthal wavenumber $m$ plotted against radii of outer Lindblad resonances for f modes and low-order g modes with predominantly even equatorial parity (even $\ell-m$), calculated from our fiducial model both with (black plus signs) and without (filled circles) the effects of Saturn's zonal winds. From right to left, the separate tracks range from $\ell-m\sim0$ to $\ell-m\sim8$ (we only plot the sectoral $n=2$ g modes). Right: same as the left-hand panel, but for modes with odd equatorial parity (odd $\ell-m$), which can excite bending waves at outer vertical resonances. The point sizes are scaled by the logarithm of the maximum surface $\Phi'$ compared against that of the $m=3$ sectoral f mode. The black open diamonds show the observed C Ring density waves and bending waves, while the black-dashed lines delimit the boundaries of the C Ring.}
    \label{fig:mvrl}
\end{figure*}

The frequencies of the f modes in our calculations agree broadly with both the perturbative calculations of \citet{Mankovich2021}, and the majority of the waves observed in Saturn's C Ring. \autoref{fig:mvrl} plots azimuthal wavenumbers $m$ against the resonant radii at which f modes (blue points) and low-order g modes (orange/green points) with predominantly even (left) and odd (right) equatorial parity would be expected to excite density waves and bending waves, respectively. We plot the observed waves as black diamonds. Filled circles denote oscillations calculated with our purely rigidly rotating model with positive Brunt-V\"ais\"al\"a frequency extending to $r_\text{stab}=0.70R_s$, while plus signs show changes in the resonant radii due to frequency shifts from differential rotation (again assuming a radial decay depth of $d\simeq7.5\times10^3$km). We scale the sizes of all the points by $1+\text{log}_{10}(\Phi'_3/\Phi'_s)$, where $\Phi'_s$ and $\Phi'_3$ are the maximum surface gravitational perturbations for the mode in question and the $m=3$ sectoral f mode, respectively.

\autoref{fig:mvrl} indicates nearly universal agreement with the observations for $m\geq5$, which is improved for the sectoral f modes (right-most blue track in the left-hand panel) by increases in mode frequency from Saturn's zonal winds. At $m=4$, an avoided crossing with the lowest-order ($n=1$) sectoral g mode makes the choice of which mode to call the f mode arbitrary. As discussed in \autoref{sec:rig}, we identify the oscillation with the largest surface gravitational perturbation as the sectoral f mode. Regardless, in this case neither oscillation quite matches the observed density wave excited at $r\simeq81\times 10^3\text{km}$. 

This discrepancy is generic to our models. In the absence of additional observations of more $m=4$ density waves, it points toward needed refinement of our picture of Saturn's deep interior, since the low-$m$ oscillations are the most affected by interactions with g modes trapped in the stably stratified cavity. Minor discrepancies between the observed $m=2$ density waves and our calculations for the $m=2$ f mode and $\ell\sim m=2$, $n=2$ g mode further emphasize this need, although \citet{Mankovich2021} found that a sharp cut-off in buoyancy frequency can help by lessening the eigenfunction overlap between these two modes, allowing their frequencies to come closer. We have focused on models with smooth variation in composition for numerical convenience, but models with discontinuous or rapid variation may provide a more appropriate description.

\subsection{Optical depth variations and detectability}\label{sec:dtau}
To identify gravito-inertial modes potentially involved in the excitation of density waves with finely split frequencies, we estimate optical depth variations induced in the C Ring by the oscillations' gravitational perturbations, following the approach of \citet{Fuller2014a} and \citet{Fuller2014b}. \autoref{app:dtau} reviews the method of these calculations. 

\subsubsection{Comparison between rigid and differential rotation}
\begin{figure*}
    \centering
    \includegraphics[width=\columnwidth]{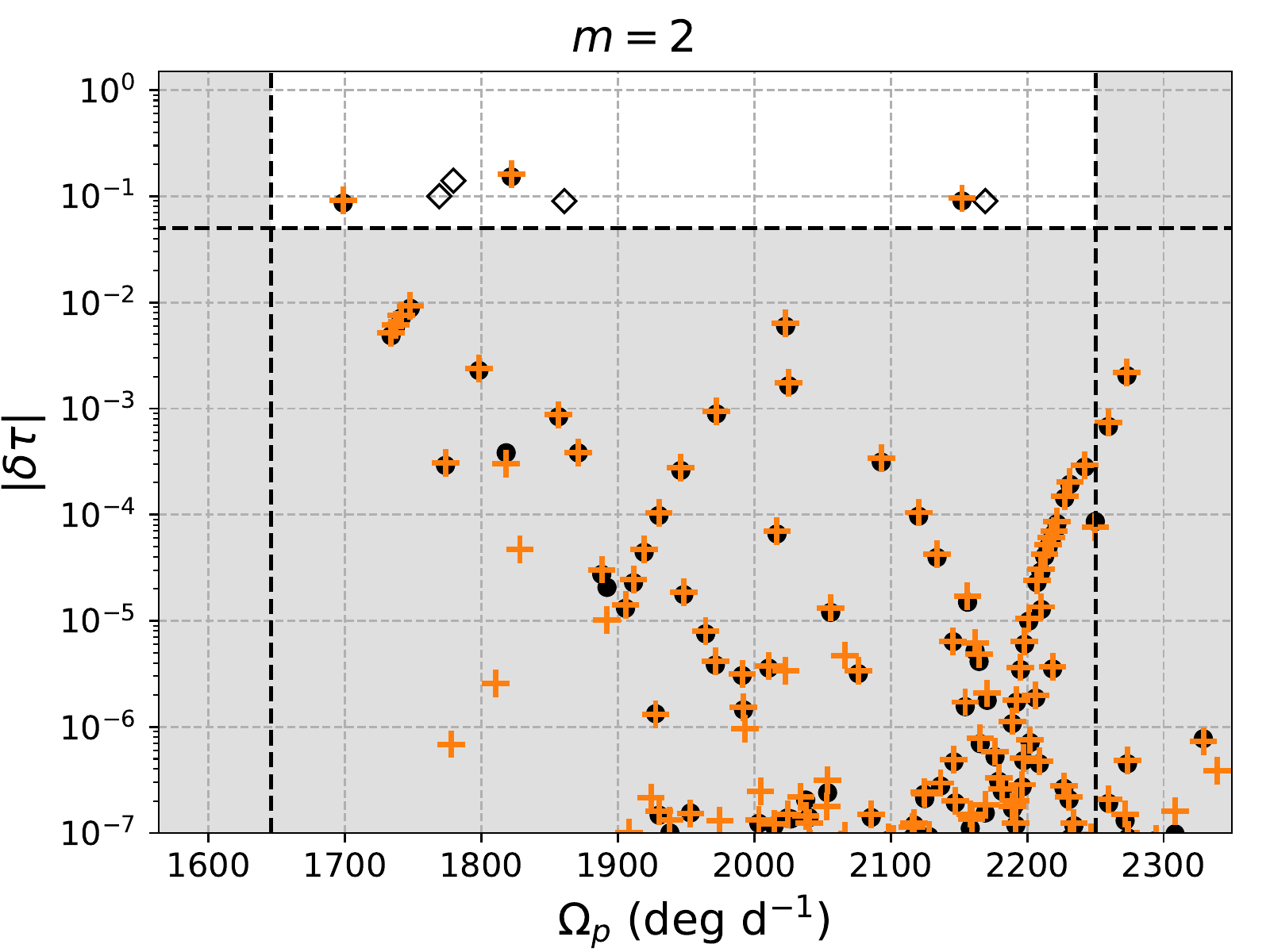}
    \includegraphics[width=\columnwidth]{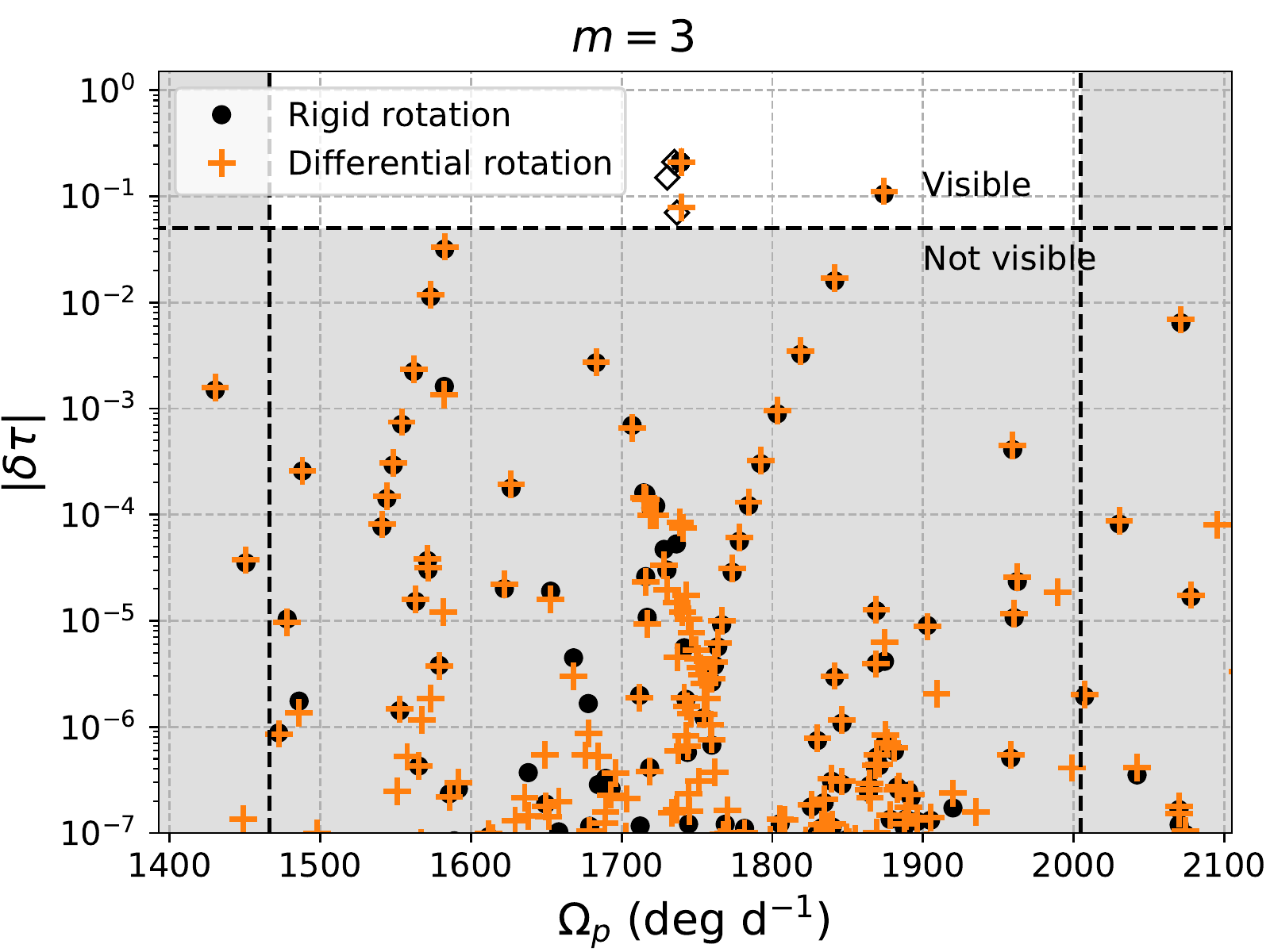}
    \caption{Predicted optical depth variations for $m=2$ (left) and $m=3$ (right) modes calculated from our model with a stably stratified region extending to $r_\text{stab}=0.70R_s$, and plotted against pattern speeds $\Omega_p=\sigma/m=\omega/m+\Omega_S$ (here $\sigma$ is the inertial-frame frequency). The black diamonds show the observed density waves, and the orange (black) points show calculations with (without) the inclusion of differential rotation from Saturn's zonal winds (assuming a decay depth of $d=0.125R_S\simeq7.5\times10^3$km). Differential rotation only appreciably shifts the frequencies of the f modes (by $\sim0.1$ per cent), but can alter the optical depth perturbations predicted for g modes.}
    \label{fig:rst68_dtau}
\end{figure*}

\autoref{fig:rst68_dtau} plots optical depth variations as a function of pattern speed $\Omega_p=\sigma/m$ (where $\sigma=\omega+m\Omega_S$ is the inertial-frame frequency) for $m=2$ (left) and $m=3$ (right) oscillations calculated for our fiducial ($r_\text{stab}=0.70R_s$) model, both with (orange) and without (black) the effects of Saturn's zonal winds assuming a radial decay depth of $d\simeq7.5\times 10^3\text{km}$. Points falling in the un-shaded region predict visible density waves, while those falling in the gray-shaded regions would be expected to produce optical depth variations too small to be observable, or resonances lying outside the C Ring. The black diamonds again correspond to the observed density waves.

The majority of modes show only minor changes in predicted $|\delta\tau|$ due to differential rotation. On the other hand, mixing between the even-parity sectoral f modes and odd-parity g modes produces a host of predicted optical depth values $|\delta\tau|\lesssim10^{-5}$ with no identifiable counterparts from the calculation with rigid rotation, since in the absence of an antisymmetric wind, odd-parity modes calculated with purely rigid rotation have $\Phi'=0$ at the equator (and therefore cannot drive density waves). Additionally, the asymmetric avoided crossing shown in \autoref{fig:rst68_avcross} produces two finely split, ostensibly visible waves close to the observed $m=3$ triplet with $\Omega_p\simeq 1736.7,$ $1735.0,$ and $1730.3\text{deg d}^{-1}$ \citep{Hedman2013}. 

Very high-$\ell$, even-parity modes with frequencies close to the $m=3$ f mode (such as the rosette mode shown in the middle-right panels of \autoref{fig:rosetteKE} and \autoref{fig:rosetteSpec}) do show $|\delta\tau|$ enhancements in some of our calculations with differential rotation (depending on the model and wind decay depth). However, we find that modes with solutions involving $\ell\gtrsim m+30$ require unreasonably small frequency separations from the sectoral f mode to produce observable optical depth variations. We note that more detailed treatments of differential rotation involving a fully self-consistent modification of the background pressure and density by the zonal winds may produce stronger coupling between the f modes and high-$\ell$ g modes. Also, our predictions of optical depth variations are sensitive to an assumption of energy equipartition between modes; any internal process that preferentially excites g modes and/or rosette modes to larger energies than f modes would be misrepresented by this assumption, imbuing the former with larger predicted $|\delta\tau|$. 

\begin{figure*}
    \centering
    \includegraphics[width=\textwidth]{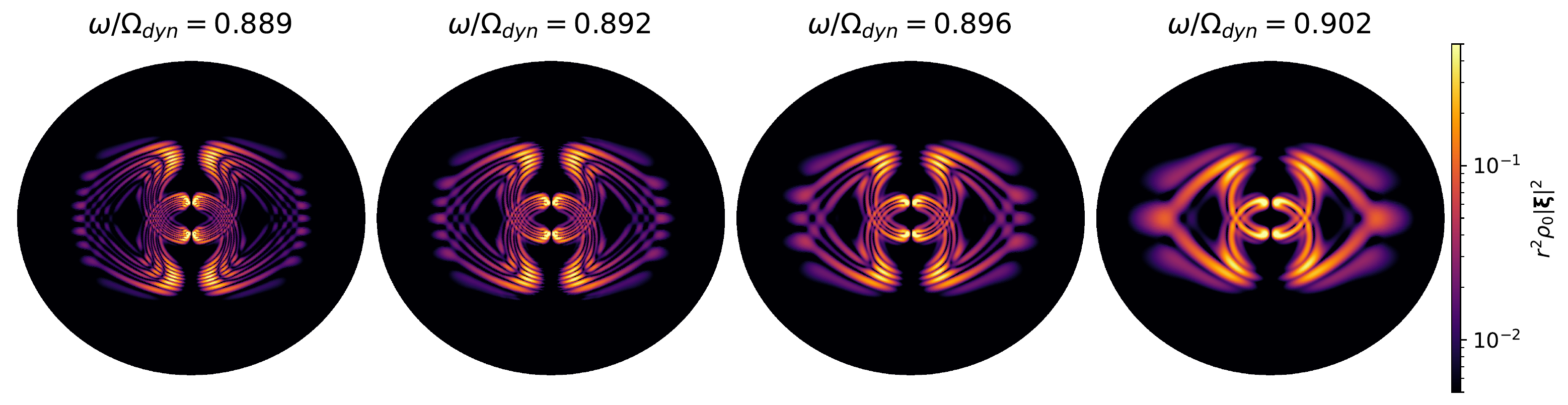}
    \caption{Meridional distributions of kinetic energy for a sequence of $m=2$ rosette modes with closely split frequencies that produce optical depth variations $|\delta\tau|\sim10^{-2}$ near $\Omega_p\sim1750\text{deg d}^{-1}$ for our model with $r_\text{stab}=0.70R_s$. This sequence produces nearly detectable optical depth variations in most of our models, and could be associated with the observed $m=2$ density waves with finely split frequencies.}
    \label{fig:rst68_rosette2}
\end{figure*}

Barring chance over-excitation by stochastic processes, such preferential excitation of g modes in the deep interior would likely require a different mechanism than those already considered for Saturn \citep{Markham2018,Wu2019}, which should operate most efficiently near the surface \citep[where mode excitation is also most efficient in the Sun; ][]{Goldreich1994}. The possibility that double-diffusive convection \citep[e.g.,][]{Moll2017} in Saturn's interior might produce alternating layers of convection and stable stratification opens the door to potentially exotic excitation mechanisms, although the formation and persistence of high-degree g modes and rosette modes with relatively short wavelengths may be more dubious in a segmented cavity \citep{Belyaev2015,Pontin2020}.

The optical depth predictions in the left-hand panel of \autoref{fig:rst68_dtau} show rough agreement with the observed $m=2$ density waves: the $m=2$ sectoral f mode produces a potentially detectable $|\delta\tau|$ close to the observed density wave with $\Omega_p\simeq1860.8\text{deg d}^{-1}$ \citep{Hedman2013}, while the sectoral $n=1$ g mode matches well with the observed density wave with $\Omega_p\simeq2169.3\text{deg d}^{-1}$ \citep{French2019}. The $\ell\sim m=n=2$ g mode produces a potentially detectable wave with a pattern speed close to, but somewhat less than the observed pair of waves at $\Omega_p\simeq1779.5\text{deg d}^{-1}$ \citep{Hedman2013} and $\Omega_p\simeq1769.2\text{deg d}^{-1}$ \citep{French2016}. As already mentioned, profiles of the Brunt-V\"ais\"al\"a frequency with a discontinuous or sharply varying cut-off in the envelope can reduce the g mode frequency spacing, and therefore shift the $n=2$ g mode closer to observations. 

\begin{figure*}
    \centering
    \includegraphics[width=\columnwidth]{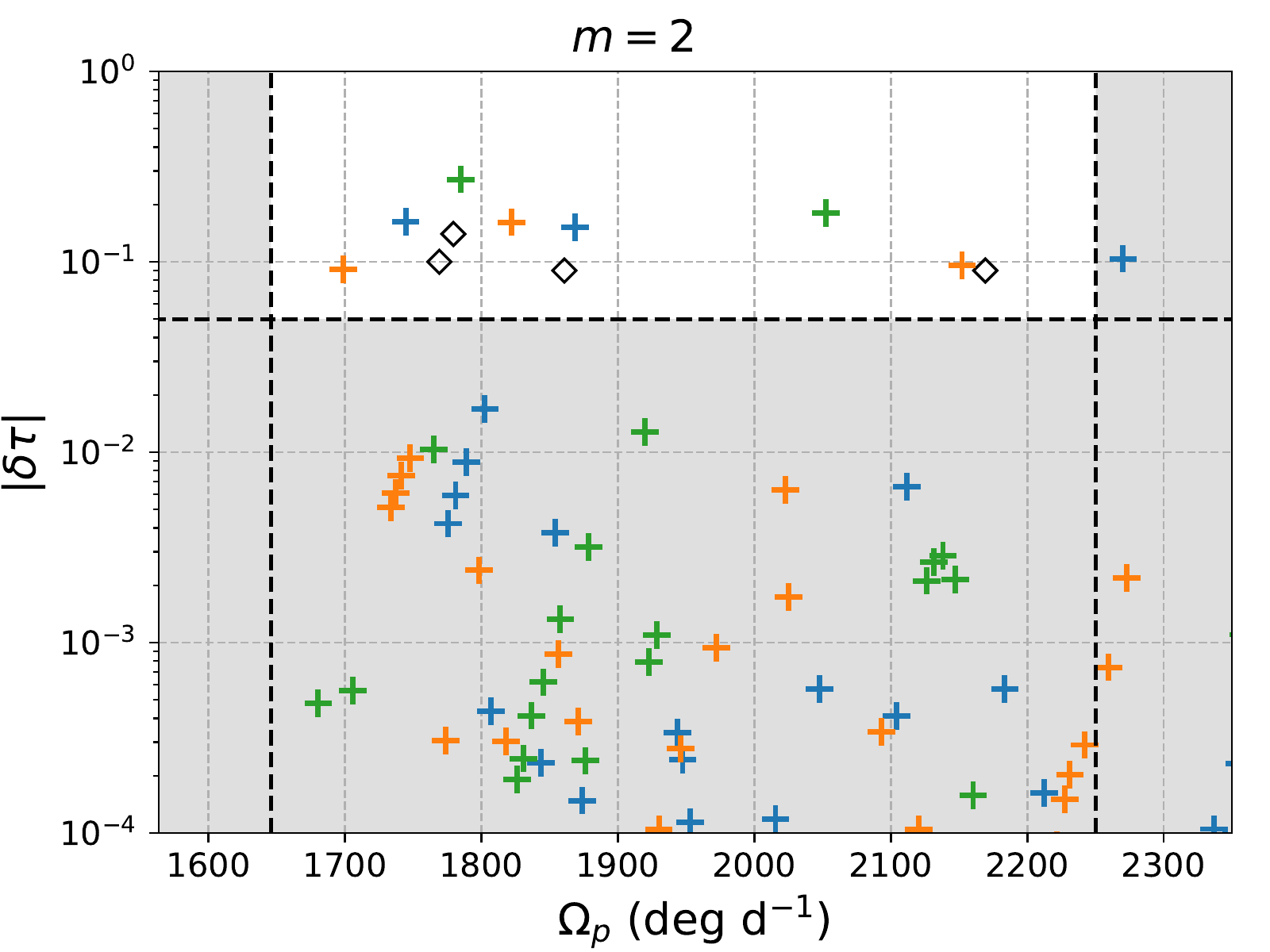}
    \includegraphics[width=\columnwidth]{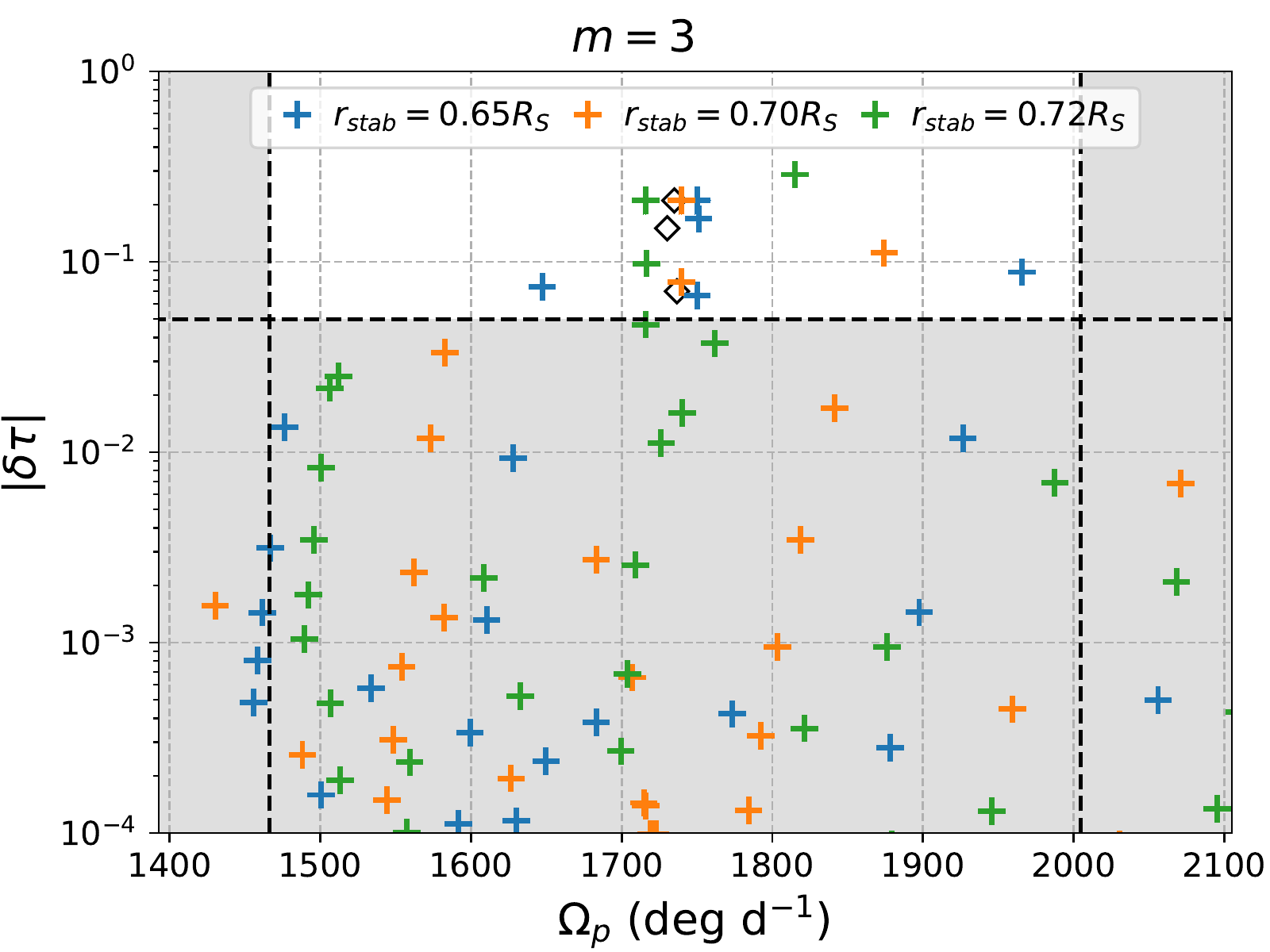}
    \caption{Same as \autoref{fig:rst68_dtau} but for all three of our models with $r_\text{stab}=0.65R_S$ (blue), $0.70R_S$ (orange) and $0.72R_S$ (green), all including differential rotation with a decay depth $d\simeq7.5\times10^3$km.}
    \label{fig:dtau_mcpr}
\end{figure*}

We also note that our calculations with most models produce a sequence of relatively low-degree rosette modes in this frequency range. For our fiducial model, these rosette modes manifest in the left-hand panel of \autoref{fig:rst68_dtau} as sequences of both black and orange points with optical variations $|\delta\tau|\sim10^{-2}$ that monotonically increase with pattern speeds $\Omega_p\gtrsim1750\text{deg d}^{-1}$. The color-plots in \autoref{fig:rst68_rosette2} show meridional distributions of kinetic energy for example rosette modes in this sequence. While the optical depth variations calculated for these modes lie just below the detectable regime, their consistent appearance (in many of our calculations with different models) in this frequency range is suggestive. With sufficient mixing with the the $n=2$ sectoral g mode, or a departure from our assumption of energy equipartition between modes (i.e., with more energy), such rosettes could be involved in exciting the finely split $m=2$ density waves. 

\subsubsection{Comparison between models}
The panels in \autoref{fig:dtau_mcpr} show $|\delta\tau|$ predictions for the three models with $r_\text{stab}=0.65R_S$, $0.70R_S$ and $0.72R_S$ considered in this paper, calculated with decay depth $d\simeq7.5\times 10^3\text{km}$. As illustrated by the blue points associated with our model with $r_\text{stab}=0.65R_S$, the three-mode interaction shown in \autoref{fig:rst65_avcross} produces a potentially detectable triplet of density waves that lie close in frequency to the observations. However, the frequency splitting of less than $\sim0.1\%$ between all three modes in this calculation is in fact too close, since the observed pattern speeds of \citep[$\Omega_p\simeq 1736.7,$ $1735.0,$ $1730.3\text{deg d}^{-1}$; ][]{Hedman2013} are split by $\sim0.1-0.35\%$. 

Additionally, while pairs of detectable waves associated with the mixing of the $m=3$ f mode with a single high-degree g mode are not difficult to produce in our calculations (due to denser spacing in frequency space for high-$\ell$ modes), engineering the overlap of such a pair with a third, relatively low-$\ell$ ($\sim5$ in this case) g mode requires significant fine-tuning. For the model with $r_\text{stab}=0.65R_S$, shifting the overall g mode spectrum to higher frequencies also shifts the $m=2$, sectoral g mode with $n=1$ away from the observed pattern speed (compare the points near $\sim2200\text{deg d}^{-1}$ in \autoref{fig:dtau_mcpr}, left). On the other hand, our model with $r_\text{stab}=0.72R_S$ significantly under-predicts the frequency of both the $n=1$ and $n=2$ sectoral g modes, the frequency for the latter falling into the sub-inertial range (which we do not consider here).

\autoref{fig:dtau_mcpr} (right) illustrates another characteristic shared by all of our calculations using the models of \citet{Mankovich2021}. Invariably, they predict that the $n=1$, $m=3$ sectoral g mode with pattern speed between $\Omega_p\sim1800\text{deg d}^{-1}$ and $1950\text{deg d}^{-1}$ (depending on the model) should excite a potentially detectable density wave. This oscillation even possesses the largest predicted $|\delta\tau|$ in our model with $r_\text{stab}=0.72R_S$, due to an enhancement of the original f mode's mode inertia via mixing with two g modes with $\ell\sim11,18$  and nearly degenerate frequencies (see \autoref{sec:pmix}). Strong frequency repulsion from the sectoral f mode likely prohibits the $n=1$ sectoral g mode from playing a role in the observed $m=3$ triplet. The fact that additional $m=3$ waves with more widely separated frequencies have not been observed may suggest that our model for Saturn's deep interior still needs refinement, or that energy equipartition may not be good assumption. 

\subsection{Simultaneous density and bending wave excitation}\label{sec:simul}
Lastly, we discuss the implications of our finding that even small antisymmetric components of a differential rotation profile can couple oscillations with even and odd equatorial parity. Parity-mixing between a sectoral f mode and an odd-parity, low-degree g mode (see \autoref{fig:rst68_avcross}) is less likely than avoided crossings involving high-degree g modes of the same parity (see \autoref{fig:rst65_avcross}); both require a close frequency degeneracy ($\lesssim0.1\%$) with the sectoral f mode, but low-$\ell$ g modes are much more sparsely spaced in frequency. Notably, though, we do find that the sectoral $m=3$ f mode and the $\ell\sim4,$ $m=3$, $n=2$ g mode have similar frequencies in many of the best-fitting models of \citet{Mankovich2021}. 

More importantly, such a coupling would have a very unique observational signature: modes with asymmetric equatorial parity could conceivably excite \emph{both} density and bending waves in Saturn's C Ring, at slightly different radii due to differences between the horizontal and vertical epicyclic frequencies associated with orbital motion. The interaction of the $m=3$ f mode with the $\ell\sim4,$ $m=3,$ $n=2$ g mode in our calculations could, for example, be confirmed by the additional detection of one or more $m=3$ bending waves at the resonant radius $r_V\simeq83\times10^3\text{km}$ (in addition to the observed density wave excitation at $r_L\simeq82.2\times10^3\text{km}$). If such parity-mixing were confirmed, the sparseness of the low-degree g mode frequency spectrum would serve as a powerful asset for seismic inference, since it would significantly reduce the number of mode interactions potentially capable of producing the observed fine splitting.

\section{Conclusions}\label{sec:conc}
We have investigated the effects of Saturn's rapid and differential rotation on the planet's normal mode oscillations, focusing on the mixing of fundamental modes (f modes) and gravito-inertial modes (g modes) likely responsible for exciting density waves with nearly degenerate frequencies in the C Ring \citep{Hedman2013,Fuller2014b}. Using a combination of non-perturbative and perturbative methods to account for Saturn's rapid bulk rotation and deep zonal winds (resp.), we have computed oscillation modes for interior models featuring wide regions of stable stratification constrained by a joint fit of Saturn's low-order seismology and gravity field \citep{Mankovich2021}. 

The stably stratified interiors in our models produce a rich spectrum of gravito-inertial oscillations (see \autoref{fig:dPh_v_om}), some adhering to the conventional understanding of g modes in non-rotating planets and stars, and others resembling so-called ``rosette modes'' (\autoref{fig:rosetteKE}; 
\autoref{fig:rosetteSpec};
\autoref{fig:rst68_rosette2}) encountered only with higher-order treatments of rapid rotation \citep{Ballot2012,Takata2013}. 

The high-degree g modes and rosette modes appear densely spaced in frequency, making their interaction with the low-$m$, sectoral f modes commonplace (\autoref{fig:rst65_avcross}; 
\autoref{fig:rst65_triplet}). We have demonstrated that such interactions can endow the otherwise-undetectable high-degree g modes with large external gravitational perturbations (relative to mode energies) ostensibly capable of exciting visible density waves with frequencies nearly degenerate to those driven by the f modes (\autoref{fig:rst68_dtau}; 
\autoref{fig:dtau_mcpr}). The avoided crossings are narrow in frequency width, however, and we infer that additional mixing with at least one lower-degree g mode may be required to reproduce an observed triplet of $m=3$ density waves.

With our perturbative treatment of differential rotation, we find that Saturn's zonal winds impact its f modes subtly but measurably. Saturn's equatorial jet most significantly alters the frequencies of high-degree sectoral modes, owing to their strict confinement to both the envelope and equatorial latitudes (\autoref{fig:rst68_fmode_freqshift};  \autoref{fig:rst68_fmode_eigfuns}; 
\autoref{fig:mvrl}). Differential rotation in Saturn's envelope also enhances the avoided crossings between f modes and g modes at low azimuthal wavenumbers $m$.

Interestingly, we find that including a realistically small antisymmetric component of the differential rotation profile leads to weak coupling between modes with different equatorial parity (\autoref{fig:rst68_avcross}). The weakness of this coupling makes parity-mixing a relatively unlikely explanation for observed fine-frequency-splitting between density waves in the C Ring. If confirmed, however, the rotational mixing between a sectoral f mode and an odd-parity g mode would provide a powerful constraint on Saturn's deep interior, since such mixing would likely be limited to a sparse spectrum of low-degree g modes. Such an interaction could be confirmed by observations of simultaneous density and bending wave excitation at Lindblad and vertical resonances with identical pattern speeds.

\acknowledgments{
The authors thank the two anonymous referees who reviewed this work, both of whom provided very thorough and constructive comments that significantly improved the quality of the paper. J. W. D. is supported by the Natural Sciences and Engineering Research Council of Canada (NSERC), [funding
reference $\#$CITA 490888-16]
}

\bibliography{saturn_diff}{}
\bibliographystyle{aasjournal}



\appendix
\allowdisplaybreaks
\section{Non-perturbative numerical method}\label{app:npert}
This appendix provides a description of the non-perturbative numerical method we use to compute the oscillations of our rigidly rotating models for Saturn. We follow closely the approach of \citet{Reese2006,Reese2009,Reese2013}, and refer the reader to these works for more details.

\subsection{Coordinate system}\label{sec:coord}
Following \citet{Reese2006,Reese2009,Reese2013}, we employ a non-orthogonal coordinate system $(\zeta,\theta,\phi)$ that matches the oblate surface of a rapidly rotating fluid body \citep{Bonazzola1998}. In units with equatorial radius $R_S=1$, $\zeta$ is a quasi-radial coordinate related to spherical radius $r$ by
\begin{equation}
    r(\zeta,\theta)=
    \begin{cases}
        (1-\epsilon)\zeta
        +\frac{1}{2}(5\zeta^3-3\zeta^5)(r_s-1+\epsilon),
        & \zeta\in[0,1],
    \\
        2\epsilon
        +(1-\epsilon)\zeta
        +(2\zeta^3-9\zeta^2+12\zeta-4)(r_s-1-\epsilon),
        & \zeta\in[1,2],
    \end{cases}
\end{equation}
where $r_s=r_s(\theta)$ defines the surface of the oblate planet, and $\epsilon=1-r_s(0)/r_s(\pi/2)$
gives the rotational flattening due to centrifugal acceleration. Denoting partial differentiation by subscripts, this coordinate system has the metric tensor
\begin{equation}
    g_{ij}
    =\left[
        \begin{matrix}
            r_\zeta^2 & r_\zeta r_\theta & 0 \\
            r_\zeta r_\theta & r^2+r_\theta^2 & 0 \\
            0 & 0 & r^2\sin^2\theta
        \end{matrix}
    \right],
\end{equation}
the inverse metric tensor
\begin{equation}
    g^{ij}
    =\left[
        \begin{matrix}
            \dfrac{r^2+r_\theta^2}{r^2r_\zeta^2} &
            \dfrac{-r_\theta}{r^2r_\zeta} &
            0 \\
            \dfrac{-r_\theta}{r^2r_\zeta} &
            \dfrac{1}{r^2} &
            0 \\
            0 & 0 & \dfrac{1}{r^2\sin^2\theta}
        \end{matrix}
    \right],
\end{equation}
and hence the Jacobian $J=\sqrt{\text{det}g_{ij}}=r_\zeta r^2\sin\theta$ \citep[e.g.,][]{Rieutord2016}. We note that, as defined, $\zeta$ need only coincide with isobars at the surface of the planetary model.

\subsection{Linearized Equations}
The linearized Equations \eqref{eq:lineq1}-\eqref{eq:lineq4} can be written in tensor notation for an arbitrary curvilinear coordinate system as
\begin{align}\label{eq:teq1}
    \text{i}\omega v^i
    &=g^{ij}\left[
        2J\epsilon_{jkl}\Omega^kv^l
        -G_jb
        +(\partial_j+\partial_j\ln\rho_0)h
        +\partial_j\Phi'
    \right],
\\
    \text{i}\omega b
    &=(J\rho_0)^{-1}\partial_j(J\rho_0v^j),
\\
    \text{i}\omega(h-c_A^2b)
    &=(G_j - c_A^2\partial_j\ln\rho_0)v^j,
\\\label{eq:teq4}
    J^{-1}\partial_j(Jg^{jk}\partial_k\Phi')&=4\pi G\rho_0 b,
\end{align}
where lower (upper) indices denote covariance (contravariance), $\epsilon_{jkl}$ is the Levi-Civita symbol, and we employ the Einstein summation convention. Like \citet{Reese2006}, we trade the contravariant vector components $v^i$ for variables $u^\zeta,u^\theta,u^\phi$ defined by 
\begin{equation}
    v^\zeta=\left(\frac{\zeta^2}{r^2r_\zeta}\right)u^\zeta,
    \hspace{2em}
    v^\theta=\left(\frac{\zeta }{r^2r_\zeta}\right)u^\theta,
    \hspace{2em}
    v^\phi=\left(\frac{\zeta }
    {r^2r_\zeta\sin\theta}\right)u^\phi.
\end{equation}
It is useful to note that the $u^i$ (not to be confused with components of the general velocity field ${\bf u}$ referenced in \autoref{eq:EoM0}) are in turn related to vector components $v_r,v_\theta,v_\phi$ associated with the unit spherical basis by \citep[e.g.,][]{Reese2013}
\begin{equation}
    v_r=\left(\frac{\zeta^2}{r^2}\right)u^\zeta + \left(\frac{\zeta r_\theta}{r^2r_\zeta}\right)u^\theta,
    \hspace{2em}
    v_\theta=\left(\frac{\zeta }{rr_\zeta}\right)u^\theta,
    \hspace{2em}
    v_\phi=\left(\frac{\zeta}{rr_\zeta}\right)u^\phi.
\end{equation}
Substituting these variables, and geometric factors from expressions from \autoref{sec:coord}, Equations \eqref{eq:teq1}-\eqref{eq:teq4} can be taken in linear combination to produce the relations
\begin{align}\label{eq:ueq1}
    -\text{i}\omega 
    \left[
        \left(\frac{\zeta^2r_\zeta}{r^2}\right)u^\zeta
        +\left(\frac{\zeta r_\theta}{r^2}\right)u^\theta
    \right]
    &=2\Omega\left(\frac{\zeta s}{r}\right)u^\phi
    +G_\zeta b
    -\rho_0^{-1}\partial_\zeta(\rho_0h)
    -\partial_\zeta\Phi',
\\
    -\text{i}\omega
    \left[
        \left(\frac{\zeta^2r_\theta}{r^2}\right)u^\zeta
        +\zeta\Theta_L u^\theta
    \right]
    &=2\Omega\zeta \Theta_Ru^\phi
    +G_\theta b
    -\rho_0^{-1}\partial_\theta(\rho_0h)
    -\partial_\theta\Phi',
\\
    -\text{i}\omega \left(\frac{\zeta}{r_\zeta}\right)u^\phi
    &=2\Omega\left(\frac{\zeta^2s}{r}\right)u^\zeta
    -2\Omega\zeta \Theta_R u^\theta
    -D_\phi\left(h+\Phi'\right),
\\
    -\text{i}\omega\rho_0b
    &=-\left(\frac{1}{ r^2r_\zeta }\right)\left[ 
        \partial_\zeta(\zeta^2\rho_0u^\zeta)
        +\left(\frac{\zeta}{s}\right)\partial_\theta(s \rho_0u^\theta)
        +\rho_0\zeta D_\phi u^\phi
    \right],
\\
    -\text{i}\omega(h-c_A^2b)
    &=-\left(\frac{\zeta}{r^2r_\zeta}\right)
    \left(
        \zeta A_\zeta u^\zeta
        +A_\theta u^\theta
    \right),
\\\label{eq:ueq6}
    0&=4\pi Gr^2\rho_0b - \left[
        \frac{(r^2+r_\theta^2)}{r_\zeta^2}
        \partial^2_{\zeta\zeta}
        +c_\zeta\partial_\zeta
        -\left(\frac{2r_\theta}{r_\zeta}\right)\partial^2_{\zeta\theta}
        +\Delta_{\theta\phi}
    \right]\Phi',
\end{align}
where we have defined, for convenience,
\begin{equation}
    \mu:=\cos\theta,
    \hspace{2em}
    s:=\sin\theta,
    \hspace{2em}
    D_\phi:=s^{-1}\partial_\phi,
    \hspace{2em}
    \Theta_L:=\frac{r^2+r_\theta^2}{r^2r_\zeta},
    \hspace{2em}
    \Theta_R:=\frac{r\mu +r_\theta s}{rr_\zeta},
\end{equation}
\begin{equation}
    A_\zeta:=G_\zeta -c_A^2\partial_\zeta\ln\rho_0,
    \hspace{2em}
    A_\theta:=G_\theta-c_A^2\partial_\theta\ln\rho_0,
\end{equation}
\begin{equation}
    c_\zeta
    :=\frac{1}{r_\zeta^3}
    \left[ 
        2r r_\zeta^2
        +2r_\theta r_\zeta r_{\zeta\theta}
        -(r^2+r_\theta^2)r_{\zeta\zeta}
        -r_\zeta^2r_{\theta\theta}
        -\cot\theta r_\theta r_\zeta^2
    \right],
    \hspace{2em}
    \Delta_{\theta\phi}
    :=\partial^2_{\theta\theta}
    +\cot\theta\partial_\theta
    +s^{-2}\partial^2_{\phi\phi}.
\end{equation}
To non-dimensionalize, we adopt the scalings $G=M_S=R_S=1$. Again writing $\Omega_\text{dyn}=\sqrt{GM_S/R_S^3}$, introducing the dimensionless variables 
\begin{equation}
    r=R_S\tilde{r},
    \hspace{0.75em}
    \zeta=R_S\tilde{\zeta},
    \hspace{0.75em}
    u^i=\left(R_S\Omega_\text{dyn}\right)\tilde{u}^i,
    \hspace{0.75em}
    \omega=\Omega_\text{dyn}\tilde{\omega},
    \hspace{0.75em}
    \rho=\left(\frac{M_S}{R_S^3}\right)\tilde{\rho},
    \hspace{0.75em}
    h=\left(\frac{GM_S}{R_S}\right)\tilde{h},
    \hspace{0.75em}
    \Phi'=\left(\frac{GM_S}{R_S}\right)\tilde{\Phi}',
\end{equation}
and immediately suppressing tildes leaves Equations \eqref{eq:ueq1}-\eqref{eq:ueq6} unaltered, save for eliminating the factor of $G$ from Poisson's equation (and altering the numerical values for the background variables $\rho_0,$ $P_0$, etc.).

\subsection{Expansion and projection}
We expand the perturbations of a fixed azimuthal wavenumber $m$ as
\begin{align}
    u^\zeta(\zeta,\theta,\phi)
    &=\sum_{l'=l_\text{min}}^\infty u^{l'}(\zeta)Y_{l'}^m ,
\\
    u^\theta(\zeta,\theta,\phi)
    &=\sum_{l'=l_\text{min}}^\infty 
    [v^{l'}(\zeta)\partial_\theta Y_{l'}^m (\theta,\phi)
    +w^{l_w'}(\zeta)D_\phi Y_{l_w'}^m(\theta,\phi)],
\\
    u^\phi(\zeta,\theta,\phi)
    &=\sum_{l'=l_\text{min}}^\infty 
    [v^{l'}(\zeta)D_\phi Y_{l'}^m (\theta,\phi)
    - w^{l_w'}(\zeta)\partial_\theta Y_{l_w'}^m(\theta,\phi)],
\\\
    b(\zeta,\theta,\phi)
    &=\sum_{l'=l_\text{min}}^\infty b^{l'}(\zeta)Y_{l'}^m(\theta,\phi),
\\
    h(\zeta,\theta,\phi)
    &=\sum_{l'=l_\text{min}}^\infty h^{l'}(\zeta)Y_{l'}^m(\theta,\phi),
\\
    \Phi'(\zeta,\theta,\phi)
    &=\sum_{l'=l_\text{min}}^\infty \psi^{l'}(\zeta)Y_{l'}^m(\theta,\phi),
\end{align}
where $Y_l^m(\theta,\phi)=\tilde{P}^m_l(\mu)e^{\text{i} m \phi}$ are orthonormal spherical harmonics with associated Legendre polynomials $\tilde{P}^m_l$ scaled to satisfy
\begin{equation}
    \int_0^{2\pi}\text{d}\phi\int_{-1}^1
    Y_l^{m*} Y_{l}^{m}\text{d}\mu
    =1.
\end{equation}
Here we set $\ell_\text{min}=m,\ell_w=\ell+1$ for calculations of modes with even equatorial parity, and $\ell_\text{min}=m+1,\ell_w=\ell-1$ for odd-parity modes. In practice, the formally infinite series must be truncated at a finite $\ell_\text{max}.$

Inserting these expansions into the linearized partial differential equations, an infinite set of coupled ordinary equations can be derived by projecting onto an arbitrary harmonic degree $\ell$; in our case, taking appropriate linear combinations of the equations multiplied by $Y_l^{m*},$ $\partial_\theta Y_l^{m*}$ and $D_\phi Y_l^{m*}$ and integrating over all solid angles \citep[as described in, e.g.,][]{Reese2006} produces
\begin{align}\label{eq:proj1}
    \omega& 
    \left[
        I_{ll'}\left(\frac{\zeta^2r_\zeta}{r^2}\right)\text{i} u^{l'}
        +J_{ll'}\left(\frac{\zeta r_\theta}{r^2}\right)\text{i} v^{l'}
        +\text{i} K_{ll_w'}\left(\frac{\zeta r_\theta}{r^2}\right)w^{l_w'}
    \right]
    \\\notag&\hspace{4em}
    =2\Omega \left[ 
        \text{i} K_{ll'}\left(\frac{\zeta s}{r}\right)\text{i} v^{l'}
        +J_{ll_w'}\left(\frac{\zeta s}{r}\right)w^{l_w'}
    \right]
    -I_{ll'}\left(G_\zeta\right)b^{l'}
    +\partial_\zeta h^{l} 
    +I_{ll'}(\partial_\zeta\ln\rho_0)h^{l'}
    +\partial_\zeta\psi^l,
\\\notag
    \omega&
    \left\{
        J^*_{ll'}\left(\frac{\zeta r_\theta}{r^2}\right)\text{i} u^{l'}
        +\left[ 
            L_{ll'}\left(\Theta_L\right)
            +N_{ll'}\left(\frac{1}{r_\zeta}\right)
        \right]\text{i} v^{l'}
        +\left[
            \text{i} M_{ll_w'}\left(\Theta_L\right)
            -\text{i} M^*_{ll_w'}\left(\frac{1}{r_\zeta}\right)
        \right]w^{l_w'}
    \right\}
    \\\notag&\hspace{4em}
    =-2\Omega \left[
        \text{i} K^*_{ll'}\left(\frac{\zeta s}{r}\right)\text{i} u^{l'}
        -\text{i} (M_{ll'}-M_{ll'}^*)\left(\Theta_R\right)\text{i} v^{l'}
        -(L_{ll_w'}+N_{ll_w'})\left(\Theta_R\right)w^{l_w'}
    \right]
    \\&\hspace{5.25em}
    -J^*_{ll'}\left(\frac{G_\theta }{\zeta}\right)b^{l'}
    +(L_{ll'}+N_{ll'})\left(\frac{1}{\zeta}\right)(h^{l'}+\psi^{l'})
    +J^*_{ll'}\left(\frac{\partial_\theta\ln\rho_0}{\zeta}\right)h^{l'},
\\\notag
    \omega&
    \left\{
        \text{i} K^*_{l_w,l'}\left(\frac{\zeta r_\theta}{r^2}\right)\text{i} u^{l'}
        -\left[ 
            \text{i} M_{l_w,l'}\left(\frac{1}{r_\zeta}\right)
            -\text{i} M^*_{l_w,l'}\left(\Theta_L\right)
        \right]\text{i} v^{l'}
        -\left[
        N_{l_w,l_w'}\left(\Theta_L\right)
        +L_{l_w,l_w'}\left(\frac{1}{r_\zeta}\right)
    \right]w^{l_w'}
    \right\}
    \\\notag&\hspace{4em}
    =-2\Omega \left[ 
        J^*_{l_w,l'}\left(\frac{\zeta s}{r}\right)\text{i} u^{l'}
        +(N_{l_w,l'}+L_{l_w,l'})\left(\Theta_R\right)\text{i} v^{l'}
        +(\text{i} M_{l_w,l_w'}-\text{i} M^*_{l_w,l_w'})\left(\Theta_R\right)
        w^{l_w'}
    \right]
    \\&\hspace{5.25em}
    -\text{i} K^*_{l_w,l'}\left(\frac{G_\theta }{\zeta}\right)b^{l'}
    -(\text{i} M_{l_w,l'}-\text{i} M^*_{l_w,l'})
    \left(\frac{1}{\zeta}\right)(h^{l'}+\psi^{l'})
    +\text{i} K^*_{l_w,l'}\left(\frac{\partial_\theta\ln\rho_0}{\zeta}\right)h^{l'},
\\\notag
    \omega& b^l
    =-\left\{
        I_{ll'}\left(\frac{\zeta^2}{r_\zeta r^2}\right)\partial_\zeta 
        +I_{ll'}\left[\frac{\partial_\zeta(\zeta^2\rho_0)}{r_\zeta r^2\rho_0}\right]
    \right\}\text{i} u^{l'}
    \\&\hspace{5.25em}
    -\left[ 
        J_{ll'}\left(\frac{\zeta\partial_\theta\rho_0}{r_\zeta r^2\rho_0}\right)
        -I_{ll'}\left(\frac{\ell'(\ell'+1)\zeta }{r_\zeta r^2}\right)
    \right]\text{i} v^{l'}
    -\text{i} K_{ll_w'}\left(\frac{\zeta\partial_\theta\rho_0}{r_\zeta r^2\rho_0}\right)w^{l_w'},
\\
    \omega&\left[h^l-I_{ll'}(c_A^2)b^{l'}\right]
    =-I_{ll'}\left(\frac{\zeta^2A_\zeta}{r^2r_\zeta}\right)
    \text{i} u^{l'}
    -J_{ll'}\left(\frac{\zeta A_\theta}{r^2r_\zeta}\right)
    \text{i} v^{l'}
    -\text{i} K_{ll_w'}\left(\frac{\zeta A_\theta }{r^2r_\zeta}\right)
    w^{l_w'},
\\\label{eq:proj6}
    0&=-4\pi I_{ll'}(r^2\rho_0)b^{l'}
    +I_{ll'}\left[\frac{(r^2+r_\theta^2)}{r_\zeta^2}\right]
    \partial^2_{\zeta\zeta}\psi^{l'}
    +\left[
        I_{ll'}(c_\zeta)
        -J_{ll'}\left(\frac{2r_\theta}{r_\zeta}\right)
    \right]\partial_{\zeta}\psi^{l'}
    -\ell(\ell+1)\psi^l,
\end{align}
where repeated indices $\ell'$ and $\ell_w'$ denote summation, and for an arbitrary function $f(\zeta,\mu)$ we have defined the integrals
\begin{align}\label{eq:I1}
    I_{ll'}(f)(\zeta)
    &=2\pi\int_{-1}^1f(\zeta,\mu)\tilde{P}_l^m\tilde{P}_{l'}^m\text{d}\mu,
    \\
    J_{ll'}(f)(\zeta)
    &=-2\pi\int_{-1}^1f(\zeta,\mu)\sqrt{1-\mu^2}
    \tilde{P}_l^m\partial_\mu\tilde{P}_{l'}^m\text{d}\mu,
    \\
    \text{i}K_{ll'}(f)(\zeta)&=-2\pi m\int_{-1}^1f(\zeta,\mu)
    \frac{\tilde{P}_l^m\tilde{P}_{l'}^m}{\sqrt{1-\mu^2}}\text{d}\mu,
    \\
    J^*_{ll'}(f)(\zeta)&=-2\pi\int_{-1}^1f(\zeta,\mu)\sqrt{1-\mu^2}
    \partial_\mu\tilde{P}_l^m\tilde{P}_{l'}^m\text{d}\mu,
    \\
    L_{ll'}(f)(\zeta)&=2\pi\int_{-1}^1f(\zeta,\mu)(1-\mu^2)
    \partial_\mu\tilde{P}_l^m\partial_\mu\tilde{P}_{l'}^m\text{d}\mu,
    \\
    \text{i}M_{ll'}(f)(\zeta)
    &=2\pi m\int_{-1}^1f(\zeta,\mu)
    \partial_\mu\tilde{P}_l^m\tilde{P}_{l'}^m\text{d}\mu,
    \\
    \text{i}K^*_{ll'}(f)(\zeta)
    &=2\pi m\int_{-1}^1f(\zeta,\mu)
    \frac{\tilde{P}_l^m\tilde{P}_{l'}^m}{\sqrt{1-\mu^2}}\text{d}\mu,
    \\
    \text{i}M^*_{ll'}(f)(\zeta)
    &=-2\pi m\int_{-1}^1f(\zeta,\mu)
    \tilde{P}_l^m\partial_\mu\tilde{P}_{l'}^m\text{d}\mu,
    \\\label{eq:Il}
    N_{ll'}(f)(\zeta)&=2\pi m^2\int_{-1}^1f(\zeta,\mu)
    \frac{\tilde{P}_l^m\tilde{P}_{l'}^m}{(1-\mu^2)}\text{d}\mu.
\end{align}
Note that the axisymmetry of our models means that each mode is characterized by a single azimuthal wavenumber, and so the repeated indices $m$ in these expressions should not be taken to imply summation.

The truncated series of coupled ODEs \eqref{eq:proj1}-\eqref{eq:proj6} can be written as a generalized eigenvalue problem with the form 
$\text{ \bf A}\cdot\text{\bf X}=\omega\text{\bf B}\cdot\text{\bf X}.$
Here $\text{\bf X}=[x^{l_\text{min}},x^{l_\text{min}+2},...,x^l,...,x^{l_\text{max}}]^T$, and in turn 
$x^l=\text{i} u^l(\zeta),\text{i}v^l(\zeta),w^{l_w}(\zeta),
b^l(\zeta),h^l(\zeta),\psi^l(\zeta),\psi^l_{out}(\zeta)$. The spectral coefficients $u^l$ and $v^l$ are multiplied by $\text{i}$ in order to make the matrices $\text{\bf A}$ and ${\text{\bf B}}$ purely real, while $\psi_{out}^l(\zeta)$ is a spectral component associated with the gravitational perturbation in the vacuum exterior to the planet. We solve for this exterior potential so that the appropriate boundary conditions (see \autoref{sec:BC}) can be applied on a spherical surface.

\subsection{Quadratures, discretization, and collocation}
\label{sec:coll}
The method of solving for oscillation modes can be summarized as i) the numerical calculation of the integrals \eqref{eq:I1}-\eqref{eq:Il}, ii) the construction of discretized matrix representations of $\text{ \bf A}$ and $\text{ \bf B}$, and iii) the computation of eigenvalues $\omega$ and eigenvectors ${\bf X}$ using standard libraries for linear algebra \citep[we use the Scipy implementations of BLAS/LAPACK and ARPACK;][]{Scipy2020}. 

We use Gaussian quadratures \citep[e.g.,][]{Press2002} with a polar grid containing $N_\mu=257$ grid points for the Galerkin step i). For step ii) we use a pseudospectral method, in which we compute the ($\zeta$-variable) coefficients in terms of their values on a Gauss-Lobatto grid comprising the endpoints and locations of extrema for Chebyshev polynomials mapped from the domain $[-1,1]$ to $\zeta\in[0,1]$ (or $[1,2]$, for the external gravitational potential). This representation allows for the use of spectral derivative matrices (derived from an expansion in a basis of the cardinal functions associated with the Gauss-Lobatto grid) wherever $\zeta$-derivatives appear in the equations, and facilitates Clenshaw-Curtis quadratures when integrating over $\zeta$ \citep{Boyd2001}.

Discarding any modes for which $\max|x^{\ell_\text{max}}|>\max|{\bf X}|/100$, we find that the remaining solutions are generally well converged with increasing $\ell_\text{max}.$ For the majority of the calculations described in this paper, we use a Gauss-Lobatto grid with $N_\zeta=120$ grid points. We choose  $\ell_\text{max}=\ell_\text{min}+40$ for all but some of the calculations for our model with $r_\text{stab}=0.70R_S$, for which we set $\ell_\text{max}=\ell_\text{min}+80$ to include higher-degree rosette modes. Equivalent calculations using the same model but setting $\ell_\text{max}=\ell_\text{min}+40$ yield minimally different results for oscillation detectability; as noted in \autoref{sec:dtau}, very high-degree g modes (like that shown in the middle right panels of \autoref{fig:rosetteKE} and \autoref{fig:rosetteSpec}) only gain significant surface $\Phi'$ when fine-tuning places their frequencies exceedingly close to those of the f modes or low-degree g modes. Increasing $N_\zeta$ also minimally affects the modes considered in this paper.

\subsection{Boundary conditions}\label{sec:BC}
With Chebyshev collocation on a Gauss-Lobatto grid, boundary conditions can be applied through the method of boundary bordering, in which matrix rows corresponding to $\zeta=0,1,2$ are replaced with the desired boundary or interface conditions. At $\zeta=0$ we impose regularity conditions (on each spherical harmonic), requiring that each $b^l,h^l,\psi^l\propto r^l$ ($u^l,v^l,w^l\propto r^{l-1}$) as $\zeta\rightarrow r\rightarrow0$ (in practice, this is done by enforcing even/odd parity with respect to $\zeta=0$). At $\zeta=1,$ we enforce the continuity of the gravitational potential and its gradient, and apply the mechanical boundary condition of a vanishing Lagrangian pressure perturbation $\Delta P=P' + {\boldsymbol{\xi}}\cdot\nabla P_0$. Expanding and projecting (and noting that $G_\theta=0$ at the surface for our models), this boundary condition can be written as
\begin{equation}
    \omega h^l
    =-I_{ll'}\left(\frac{\zeta^2G_\zeta}{r^2 r_\zeta}\right)
    \text{i} u^{l'}.
\end{equation}
Lastly, the solution for the gravitational potential that vanishes at infinity is $\psi^\ell_\text{out}\propto r^{-(\ell+1)},$ implying the boundary condition
\begin{equation}
    \frac{1}{r_\zeta}\frac{\text{d}\psi^l_\text{out}}{\text{d}\zeta}
    +\left(\frac{\ell+1}{\zeta}\right)\psi^l_\text{out}=0,
\end{equation}
which we apply on the spherical surface $\zeta=r=2$.

\section{Optical depth variations}\label{app:dtau}
Consider the gravitational potential produced by each mode in an inertial frame external to the planet:
\begin{align}
    &\Phi_{ext}'
    =Ae^{-\text{i}\sigma t}\sum_{l'=|m|}^{l_\text{max}}\psi_\text{ext}^{l'}Y_{l'}^m,
\end{align}
where $\sigma=\omega+m\Omega_S$ is the mode's inertial-frame frequency, and $A$ its amplitude. In an equatorial ring with orbital frequency $\Omega_r$ and surface density $\Sigma_r$, the effective forcing potential for a density wave with azimuthal wavenumber $m$ is 
\begin{equation}
    \Psi(t)
    =\left(
        \frac{\text{d}}{\text{d}\ln r}
        +\frac{2m\Omega_r}
        {m\Omega_r-\sigma}
    \right)\Phi_{ext}'(r,\theta=\pi/2,\phi,t),
\end{equation}
evaluated at the Lindblad radius $r_L$ \citep{goldreich1979}. This Lindblad radius can be computed by solving the equation $(m\Omega_r - \sigma)=-\kappa_r$, where $\kappa_r$ is the horizontal epicyclic frequency. Following \citet{Mankovich2019}, we use a multipole expansion for $\Omega_r$ and $\kappa_r$, and solve for $r_L$ numerically. 

The forcing potential $\Psi$ excites spiral density waves that, in the (thoroughly justified) ``tight-winding'' limit and WKB approximation, induce gravitational perturbations with the form
\begin{equation}
    \Phi_r'\simeq-\Psi
    \left(\frac{4\pi^2G\Sigma_r}{r_L\mathcal{D}}\right)^{1/2}
    \exp\left[\frac{\text{i}r_L\mathcal{D}}{4\pi G\Sigma_r}x^2
    \right],
\end{equation}
where $x=(r-r_L)/r_L$, and $\mathcal{D}\simeq 3(m-1)\Omega_r^2|_{r_L}$ near a Lindblad resonance \citep{Cuzzi1984}. In turn, $\Phi_r'$ can be related to a surface density perturbation $\delta\Sigma$ and optical depth variation $\delta\tau=\kappa_m\delta\Sigma$ (here $\kappa_m$ is the local mass extinction coefficient) via
\begin{equation}
    \delta\Sigma=\frac{\delta\tau}{\kappa_m}
    =\frac{\text{i}}{2\pi Gr^{1/2}}
    \frac{\text{d}(r^{1/2}\Phi_r')}{\text{d}r }
    \simeq\frac{x\Psi}{2\pi }\left(\frac{\mathcal{D}}{G^3\Sigma_rr_L}\right)^{1/2}
    \exp\left[\frac{\text{i}r_L\mathcal{D}}{4\pi G\Sigma_r}x^2
    \right],
\end{equation}
such that at a given location $x$ in the ring,
\begin{equation}
    |\delta\tau|\simeq 
    A\left|
        \frac{3(m-1)\kappa_m^2\Omega_r^2|\Psi/A|^2}{4\pi^2G^3\Sigma_r r_L}
    \right|^{1/2}|x|.
\end{equation}

Like \citet{Fuller2014b}, we use this direct relationship between $\delta\tau$ and mode amplitude to calculate the amplitude $|A_3|$ of the $\ell\sim|m|=3$ f-mode that would be required to produce the optical depth variation $\delta\tau\simeq0.21$ observed for the largest amplitude $|m|=3$ density wave in Saturn's rings. The assumption of energy equipartition with $\omega^2|A|^2=\omega_3^2|A_3|^2$ then provides amplitudes for the rest of the modes (all of which are normalized ahead of time according to \autoref{eq:norm}). We set  $|r-r_L|=5\text{km}$, and adopt the nominal values  $\kappa_m=0.02\text{cm}^2\text{g}^{-1}$ and  $\Sigma_r=5\text{g}\text{cm}^{-2}$. These are roughly appropriate to the location of the $m=3$ triplet \citep{Hedman2013}, but we note that conditions vary at the locations of many of the other observed waves.

\end{document}